\documentclass[longauth]{aa} 

\usepackage{graphicx}
\usepackage{xcolor}
\usepackage{array}
\usepackage{multirow}
\usepackage{lipsum}
\usepackage[switch]{lineno}
\usepackage{float}
\usepackage{makecell}
\usepackage{ulem}
\usepackage{comment}
\usepackage{amssymb}
\usepackage{scrextend} 
\usepackage{placeins} 
\usepackage{hyperref}
\hypersetup{
    colorlinks=true,
    linkcolor=blue,
    filecolor=magenta,      
    urlcolor=blue,
    citecolor=blue,
    }

\usepackage{txfonts}

\graphicspath{{}}

\begin{document}

    \title{Detection of RS Oph with LST-1 and modelling of its HE/VHE gamma-ray emission}

   \titlerunning{Detection of RS Oph with LST-1 and modelling of its HE/VHE gamma-ray emission}

    \author{\small
    K.~Abe\inst{1} \and
    S.~Abe\inst{2} \and
    A.~Abhishek\inst{3} \and
    F.~Acero\inst{4,5} \and
    A.~Aguasca-Cabot\inst{6}$^*$ \and
    I.~Agudo\inst{7} \and
    C.~Alispach\inst{8} \and
    N.~Alvarez~Crespo\inst{9} \and
    D.~Ambrosino\inst{10} \and
    L.~A.~Antonelli\inst{11}
% ...
    }
    \institute{
    Department of Physics, Tokai University, 4-1-1, Kita-Kaname, Hiratsuka, Kanagawa 259-1292, Japan
    \and Institute for Cosmic Ray Research, University of Tokyo, 5-1-5, Kashiwa-no-ha, Kashiwa, Chiba 277-8582, Japan
    \and INFN and Università degli Studi di Siena, Dipartimento di Scienze Fisiche, della Terra e dell'Ambiente (DSFTA), Sezione di Fisica, Via Roma 56, 53100 Siena, Italy
    \and Université Paris-Saclay, Université Paris Cité, CEA, CNRS, AIM, F-91191 Gif-sur-Yvette Cedex, France
    \and FSLAC IRL 2009, CNRS/IAC, La Laguna, Tenerife, Spain
    \and Departament de Física Quàntica i Astrofísica, Institut de Ciències del Cosmos, Universitat de Barcelona, IEEC-UB, Martí i Franquès, 1, 08028, Barcelona, Spain
    \and Instituto de Astrofísica de Andalucía-CSIC, Glorieta de la Astronomía s/n, 18008, Granada, Spain
    \and Department of Astronomy, University of Geneva, Chemin d'Ecogia 16, CH-1290 Versoix, Switzerland
    \and IPARCOS-UCM, Instituto de Física de Partículas y del Cosmos, and EMFTEL Department, Universidad Complutense de Madrid, Plaza de Ciencias, 1. Ciudad Universitaria, 28040 Madrid, Spain
    \and INFN Sezione di Napoli, Via Cintia, ed. G, 80126 Napoli, Italy
    \and INAF - Osservatorio Astronomico di Roma, Via di Frascati 33, 00040, Monteporzio Catone, Italy
% ...
}
    
   \date{Received 01 October 2024 / Accepted 15 January 2025}

   \offprints{lst-contact@cta-observatory.org, \\$^{*}$Corresponding authors}

  \abstract
   {The recurrent nova RS~Ophiuchi (RS~Oph) underwent a thermonuclear eruption in August 2021. In this event, RS~Oph was detected by the High Energy Stereoscopic System (\mbox{H.E.S.S.}), the Major Atmospheric Gamma Imaging Cherenkov (MAGIC), and the first Large-Sized Telescope (\mbox{LST-1}) of the future Cherenkov Telescope Array Observatory (CTAO) at very-high gamma-ray energies above $100\,\rm{GeV}$. This means that novae are a new class of very-high-energy (VHE) gamma-ray emitters.}
   {We report the analysis of the RS~Oph observations with \mbox{LST-1}. We constrain the particle population that causes the observed emission in hadronic and leptonic scenarios. Additionally, we study the prospects of detecting further novae using \mbox{LST-1} and the upcoming LST array of CTAO-North.}
   {We conducted target-of-opportunity observations with \mbox{LST-1} from the first day of this nova event. The data were analysed in the framework of \texttt{cta-lstchain} and \texttt{Gammapy}, the official CTAO-LST reconstruction and analysis packages. One-zone hadronic and leptonic models were considered to model the gamma-ray emission of RS~Oph using the spectral information from \textit{Fermi}-LAT and \mbox{LST-1}, together with public data from the MAGIC and \mbox{H.E.S.S.} telescopes.}
   {RS~Oph was detected at $6.6\sigma$ with \mbox{LST-1} in the first 6.35 hours of observations following the eruption. The hadronic scenario is preferred over the leptonic scenario considering a proton energy spectrum with a power-law model with an exponential cutoff whose position increases from $(0.26\pm 0.08)\,\mathrm{TeV}$ on day 1 up to $(1.6\pm 0.6)\,\mathrm{TeV}$ on day 4 after the eruption. The deep sensitivity and low energy threshold of the LST-1/LST array will allow us to detect faint novae and increase their discovery rate. 
   }
   {}


   \maketitle
%

\section{Introduction}\par
Novae are thermonuclear runaway explosions on the surface of white dwarfs in binary systems \citep{2021ARA&A..59..391C}. Since the addition of novae as a new source class that emits in the high-energy gamma-ray sky (HE; $E > 100\,\rm{MeV}$; \citealt{FermiNovae2014Sci}), novae have generated interest in the very-high-energy (VHE; $E>100\,\rm{GeV}$) gamma-ray domain for their potential to accelerate particles to TeV energies efficiently \citep[e.g.,][]{2016MNRAS.457.1786M}. Novae also provide an excellent opportunity to study particle acceleration on fast shock-evolution timescales, as well as in abundant systems within the Milky Way \citep[with a Galactic nova rate estimated from different studies between 26--50\,yr$^{-1}$;][]{2023MNRAS.523.3555Z,2022ApJ...937...64K,2022ApJ...936..117R,2021ApJ...912...19D,2017ApJ...834..196S}. Nevertheless, nova observations did not succeed in a detection at VHE gamma rays \citep{Aliu_2012,2015A&A_Ahnen,Albert_2022} until the 2021 eruption of RS~Ophiuchi \citep[RS~Oph;][]{2022_RSOphHESS,2022_RSOphMAGIC}.

The source RS~Oph is a well-known binary system that experiences recurrent nova explosions that range from 8.6 up to 26.6 years (see \citealt{2010ApJS..187..275S} for a review). 
These explosions result from the high mass-accretion rate onto the massive white dwarf driven by the giant companion star. The accretion process in RS~Oph is unclear, but the donor star may overfill its Roche lobe \citep{2017MNRAS.464.2784S,2009ApJ...697..721S,2016MNRAS.457..822B}. RS~Oph is characterised as an embedded nova because its eruptions occur immersed in the dense wind of the post-main-sequence companion star \citep[M0 III;][]{1999AA...344..177A}. The red giant in these systems produces dense circumbinary material, likely concentrated in the orbital plane. Relativistic particles accelerated via diffusive acceleration in expanding shocks are thought to interact with the dense circumbinary material, generating HE gamma rays \citep{2010Sci...329..817A,2012BaltA..21...62H,2013A&A_Martin}. 

The first nova detection at HE gamma rays was reported by the \textit{Fermi}-LAT Collaboration in the symbiotic system V407 Cyg \citep[][]{2010Sci...329..817A}. Subsequent nova detections at HE gamma rays from several binary systems and, in particular, those with main-sequence companion stars \citep[known as classical novae;][]{FermiNovae2014Sci}, likely indicate that particle acceleration is an intrinsic phenomenon in nova systems \citep{10.1093/mnras/stw2776}. The detection of classical novae, which do not exhibit a dense circumbinary material, suggests that internal shocks between several outflows can act as another mechanism to accelerate particles in novae \citep{10.1093/mnras/stu844,2018A&A_Martin,2021ARA&A..59..391C}.

Most novae that were detected at HE gamma rays are not embedded in a dense environment. When novae erupt, the detected HE emission presents a similar curved spectral shape regardless of their type \citep{FermiNovae2014Sci}. However, the HE luminosity and duration differ and vary depending on the systems \citep[e.g.,][]{FermiNovae2014Sci, 2016ApJ...826..142C, 2023MNRAS.521.5453S}. Early studies with a limited sample of classical novae suggested an inverse relation between the HE gamma-ray luminosity and its duration \citep{2016ApJ...826..142C}. However, with a larger sample, this relation may no longer hold. Novae seem to emit for a longer time at HE gamma rays, however, when they take longer to decline from the optical maximum \citep{Albert_2022}. Deep gamma-ray observations with multi-wavelength data are required to determine whether the physical differences between the two nova types also reflect a different HE--VHE gamma-ray emission.

During the 2021 nova event of RS~Oph, the High Energy Stereoscopic System (\mbox{H.E.S.S.}) and the Major Atmospheric Gamma Imaging Cherenkov (MAGIC) telescope facilities detected RS~Oph \citep{2022_RSOphHESS, 2022_RSOphMAGIC}. RS~Oph is the first nova that was detected at VHE gamma rays and the first confirmation of particle acceleration up to TeV energies in embedded novae. However, the exact acceleration and radiation mechanisms remain unclear, although the favoured explanation for the VHE gamma-ray emission is the hadronic scenario \citep{2018A&A_Martin, 2022_RSOphMAGIC, 2022_RSOphHESS, 2022PhRvD.106j3011Z, Diesing_2023, De_Sarkar_2023}.

The Large-Sized Telescope prototype (\mbox{LST-1}) of the upcoming Cherenkov Telescope Array Observatory (CTAO) observed RS~Oph during the nova phase. RS~Oph is the first transient source detected with \mbox{LST-1} during its commissioning phase. In this work, we report the spectral analysis and modelling of RS~Oph using \mbox{LST-1} and \textit{Fermi}-LAT observations and exploiting published gamma-ray data by \mbox{H.E.S.S.} and MAGIC. In Sect.~\ref{sect:obs_data_analysis} the observations and the analyses of \mbox{LST-1} and contemporaneous \textit{Fermi}-LAT data with \mbox{LST-1} are described. In Sect.~\ref{sect:modelling} we introduce the model we used to characterise the gamma-ray emission of RS~Oph. In Sect.~\ref{sect:results} the results from the \mbox{LST-1} data analysis and the modelling using the \textit{Fermi}-LAT, \mbox{LST-1}, MAGIC, and \mbox{H.E.S.S.} spectral information are presented. In Sect.~\ref{sect:discussion_outlook} we discuss the results and future expectations for nova detections with the array of LSTs of CTAO. Concluding remarks are provided in Sect.~\ref{sect:conclusion}.

\section{Observations and data analysis}
\label{sect:obs_data_analysis}
In this section, we describe the observation campaign conducted on RS~Oph with \mbox{LST-1} and the analysis procedures we used in this work (Sect.~\ref{sect:LST1analysis}). We also analyse \textit{Fermi}-LAT data that were obtained contemporaneously with the \mbox{LST-1} observations (Sect.~\ref{sect:FermiLATanalysis}).

\subsection{\mbox{LST-1}}\par
\label{sect:LST1analysis}

The LST is the largest telescope type of the upcoming CTAO. \mbox{LST-1} is the first out of four LSTs that will constitute the LST array of CTAO in the Northern Hemisphere \citep[CTAO-North;][]{2019APh...111...35A}. \mbox{LST-1} is located in the Observatorio del Roque de los Muchachos on the Canary island of La Palma, Spain. Equipped with a 23 m diameter mirror dish, LSTs have a large light-collection area with a camera consisting of high quantum-efficiency photomultiplier tubes. The trigger threshold is optimised to achieve the lowest gamma-ray energy threshold, which is about 20 GeV before the cleaning stage. 
This makes LSTs ideal telescopes for the observation of gamma-ray sources at energies from tens to hundreds of GeV \citep{LSTPerformance}.

\mbox{LST-1} started follow-up observations of RS~Oph based on its detection with \textit{Fermi}-LAT \citep{Fermi_Atel} at HE gamma rays and its bright emission in optical wavelengths. The first \mbox{LST-1} observation was recorded on August 9 (MJD 59435.90), about one day after the optical trigger. \mbox{LST-1} observed RS~Oph for several days between August 9 and September 2 (MJD 59459.91). In this work, we analyse \mbox{LST-1} data in good atmospheric\footnote{Atmospheric transmission at 9\,km above 80\%.} and dark or low-moonlight observing conditions\footnote{Observations with the Moon below the horizon or maximum diffuse night-sky-background level below 2.3 photoelectrons.} during the observation campaign (see Table~\ref{tab:obs_table}).

\begin{table}
    \caption{
    Observation campaign with \mbox{LST-1}. 
    }
    \begin{center}
        \footnotesize
        \label{tab:obs_table}
        \begin{tabular}{cccc}
        \hline \hline
        \noalign{\smallskip}
        Start date &  $t-t_0$ &  Effective time after data selection \\
        $\rm{[MJD]}$ & [d] & [h] \\
        \noalign{\smallskip}
        \hline
        \noalign{\smallskip}
         ~59435.90$^*$ &  ~\,0.97 & ~\,1.43\\
         ~59436.90$^*$ &  ~\,1.97 & ~\,2.68\\
         ~59438.90$^*$ &  ~\,3.97 & ~\,2.24\\
         59455.94 &  21.01 & ~\,0.97\\
         59456.90 &  21.97 & ~\,1.52\\
         59458.98 &  24.05 & ~\,0.32\\
         59459.91 &  24.98 & ~\,1.27\\
        \noalign{\smallskip}
        \hline
        \end{tabular}
    \end{center}
    \tablefoot{For each observation day, we list the starting date, the starting time offset with respect to the optical trigger \citep[$t_0=\rm{MJD}\,59434.93$;][]{VSNET_alert}, and the effective observation time after the data selection. We mark the dates for which daily spectral analyses were performed with an asterisk.}
\end{table}

The \mbox{LST-1} observations were performed in wobble mode with an offset of 0.4\degr\ \citep{FOMIN1994137}. The \mbox{LST-1} data were reduced from raw signal waveforms to a list of gamma-like events using the software package \verb|cta-lstchain| \citep{2022ASPC..532..357L} following the \mbox{LST-1} standard source-independent analysis approach for the calibration, image cleaning, and parametrisation, gamma-hadron separation, energy and direction reconstruction, and gamma-ray event selection \citep{LSTPerformance}. The classification and regression methods rely on random forest algorithms trained on Monte Carlo (MC) gammas and protons, simulated following a declination track of $-$4.13\degr\ in the sky plane, close to the RS~Oph declination\footnote{MC simulations at predefined declination values were simulated across the sky to account for the magnetic field effect on the extensive air shower development. The impact of using a slightly different declination value for the MC simulations and RS~Oph is negligible.}. The instrument response functions (IRFs) for each \mbox{LST-1} observation were produced by interpolating the IRFs calculated at each sky direction of the test MC data to the average telescope pointing direction of each observation. The open-software package \texttt{Gammapy} \citep{2023A&A...678A.157D, gammapy_v1.0} was used to obtain the scientific products from the gamma-like events, which were assigned based on event selection cuts. For instance, the event angular separation with respect to the source position ($\theta$) and gammaness\footnote{A parameter that indicates how likely it is that the event is a gamma-ray event.} parameters were used in this step. To assess the signal, the same cuts in the whole energy range were used for the above-mentioned parameters. Conversely, the event selection cuts applied to the gamma-like events used to compute the spectral energy distributions (SEDs) and light curve are energy dependent: The cut was set at the value at which 70\% of the MC testing gamma rays survive in each energy bin for the $\theta$ and gammaness parameters.

We performed a spectral analysis using control sky regions (OFF regions) located around the telescope pointing at the same offset as RS~Oph. OFF regions at angular offsets of 90\degr, 180\degr\, and 270\degr\ were used, where 0\degr\ is towards the telescope pointing and the position of RS~Oph. OFF regions were used to subtract the background. The energy threshold of the analysis, computed as the peak position of the energy distribution of the simulated gamma-ray events from a source with a power-law spectral index equal to $-$4, is $E\sim 30 \,\rm{GeV}$. 

We computed the integral fluxes on a daily basis for observations immediately after the eruption \citep[$t-t_0 < 4\, \rm{d}$, where $t_0=\rm{MJD}\,59434.93$;][]{VSNET_alert} using the spectral shape from the best-fit model for each day. Moreover, we computed integral fluxes by taking several observation days together (hereafter called joint flux; see observation days in Table~\ref{tab:obs_table}), employing the best-fit spectral model in the corresponding time period. We calculated upper limits (ULs) at the 95\% confidence level without considering the systematic uncertainty of the telescope energy scale. The error uncertainties correspond to the 1$\sigma$ statistical errors.

\subsection{\textit{Fermi}-LAT}\par 
\label{sect:FermiLATanalysis}

The first detection of RS~Oph at gamma rays was reported with \textit{Fermi}-LAT, coincident with the optical discovery \citep{Fermi_Atel}. The temporal trend in the HE band is similar to the trend in the optical band: A flat peak emission around $\sim$1\,day after the eruption, preceded by a smooth power-law increase and a subsequent power-law decay (slopes between wavelengths consistent within errors). However, the gamma-ray onset was delayed by $\sim$0.35\,day with respect to the time of the eruption and could have reached the peak at later times than in the optical by about 0.5\,day \citep{2022ApJ...935...44C}. The HE gamma-ray emission presents significant spectral curvature, which hardens as the eruption evolves in time. The preferred origin for the HE emission is hadronic, as the model effectively explains the observed emission. No leptonic model was tested in \citet{2022ApJ...935...44C}. We performed a dedicated \textit{Fermi}-LAT analysis to obtain contemporaneous gamma-ray spectra with the LST-1 observations of RS~Oph.

To analyse the \textit{Fermi}-LAT data, we considered reconstructed events between $50\, \rm{MeV}$ and $300\, \rm{GeV}$ with \verb|evclass=128| and \verb|evtype=3|. Only events with good time intervals (DATA\_QUAL>0 \&\& LAT\_CONFIG==1) coming below a zenith angle of 90\degr\ were selected. A binned analysis within a region of interest of 20\degr\ around RS~Oph was used to model the projected area of interest. We considered in the model all sources in the LAT 10-year source catalogue (4FGL-DR2) together with the Galactic diffuse and the standard isotropic background from Pass 8. All the spectral parameters of all sources within 4\degr\ to RS~Oph were let free to vary above $ 50\,\mathrm{MeV}$. No spectral differences were observed when we fixed the spectral parameters of 4FGL J1745.4$-$0753, the closest source to RS~Oph, to its catalogue value. We adopted the latest IRFs (\verb|P8R3_SOURCE_V3|) in the analysis. The data processing and analysis were performed using \verb|Fermitools| version 2.0.8 and \verb|Fermipy| version 1.0.1 \citep{2019ascl.soft05011F,2017ICRC...35..824W}.

We extracted the RS~Oph daily SED for the days on which \mbox{LST-1} observed RS~Oph. A log-parabola spectral shape was used to model the HE gamma-ray emission \citep{2022_RSOphMAGIC}. The ULs were computed at the 95\% confidence level for energy bins with test statistic ($TS$) values below 4, and the error uncertainties correspond to the 1$\sigma$ statistical errors. Additionally, when a significant flux was between two ULs, it was set to UL\footnote{This restrictive cut was set to ensure that the \textit{Fermi}-LAT significant differential fluxes were as robust as possible for the model fitting.}.

\section{Modelling}
\label{sect:modelling}

Proton-proton interactions are thought to be the mechanism for the gamma-ray emission in RS~Oph \citep{2022_RSOphMAGIC, 2022_RSOphHESS, 2022ApJ...935...44C}. However, a leptonic contribution to the observed emission cannot be discarded \citep{De_Sarkar_2023}. 

We considered the same modelling as was used in \citet{2022_RSOphMAGIC} to model the gamma-ray emission of RS~Oph. We considered hadronic and leptonic scenarios to explain its gamma-ray emission. We parametrised the particle spectrum as an exponential cutoff power-law (ECPL) model, while a broken power-law (BPL) model was also considered for the electron spectrum. The same nova parameter values for RS~Oph (e.g. distance, ejecta velocity, photosphere radius, and temperature) as were used in \citet{2022_RSOphMAGIC} were assumed in this work (see Table \ref{tab:nova_model_params_table}) because the data are simultaneous.

\begin{table}
    \caption{
     Nova parameters used to model the RS Oph gamma-ray SED. 
    }
    \begin{center}
        \footnotesize
        \label{tab:nova_model_params_table}
        \begin{tabular}{lccc}
        \hline \hline
        \noalign{\smallskip}
        Parameter  & \multicolumn{3}{c}{Values on observation day} \\
        \cline{2-4}
        \noalign{\smallskip}
           &  Day 1  &  Day 2  &  Day 4 \\
        \noalign{\smallskip}
        \hline
        \noalign{\smallskip}
          Distance [kpc] &  2.45 & 2.45 & 2.45 \\
          Photosphere radius [$R_{\rm \odot}$] & 200 & 200 & 200 \\
          Photosphere temperature [K] &  10780 & 9490 & 7680 \\
          Time after nova explosion [d] & 1 & 2 & 4 \\
          Expansion velocity [km s$^{-1}$] &  4500 & 4500 & 4500 \\
          Mass of nova ejecta [10$^{-6}\,M_{\rm \odot}$] & 1 & 1 & 1 \\
          Confinement factor\tablefootmark{a} & 0.1 & 0.1 & 0.1 \\
        \noalign{\smallskip}
        \hline
        \end{tabular}
    \end{center}
    \tablefoot{ For each observation day, we list the distance, the photosphere radius, the photosphere temperature, the time after the nova explosion, the expansion velocity, the mass of nova ejecta, and the confinement factor. The values were extracted from supplementary table 10 of \citet[][]{2022_RSOphMAGIC}.}
    \tablefoottext{a}{\footnotesize{The relative thickness of the shell of the expelled material.}}
\end{table}

The model maps the ejecta close to a thin layer for the energetic particles. The processes involved in the gamma-ray production for the hadronic scenario are the decay of neutral and charged pions, whereas the inverse-Compton (IC) process alone is considered for the leptonic scenario. Bremsstrahlung emission is negligible because the total column density is lower than the radiation length of hydrogen \citep{2022_RSOphMAGIC}. The seed of photons that dominate during the eruption comes from the photosphere, whose temperature evolves in time (see Table~\ref{tab:nova_model_params_table}). Gamma-gamma absorption is considered in the model. However, it is only relevant for the first day after the eruption \citep{2022_RSOphMAGIC}.

The SED data points at gamma rays from \textit{Fermi}-LAT, \mbox{LST-1}, MAGIC, and \mbox{H.E.S.S.} were used in the model fitting. As we combine data from multiple instruments, we adapted the model fitting to account for systematic uncertainties in the energy scale of the spectra obtained by the imaging atmospheric Cherenkov telescope (IACT) facilities. A systematic uncertainty in the energy scale between $\pm 15\%$ was considered as a nuisance parameter for each experiment on each day during the fitting process. The absolute $15\%$ maximum value was assumed based on the reported energy-scale uncertainty from MAGIC and \mbox{H.E.S.S.} \citep{ALEKSIC201676,2022_RSOphHESS}. Similar systematic uncertainties are expected for \mbox{LST-1} \citep[see Fig. 11 in][]{2023A&A...680A..66A}. Normally, energy-scale systematics are the dominant effect in IACT. This is especially relevant for soft sources such as RS~Oph. However, other types of systematics may contribute, such as uncertainties in the background normalisation, especially at lower energies for a single telescope \citep[][]{LSTPerformance}. Including systematic uncertainties in the model fitting reduces the number of degrees of freedom in the fitting process because a displacement of the SED points of the IACTs is allowed with respect to the original points. ULs are not included in the model-fitting minimisation.

The small distance between the \mbox{LST-1} and MAGIC telescopes means that the measurements with both instruments are not fully independent. Namely, the same gamma-ray shower can be registered by both. The correlation is expected to be energy-dependent and is difficult to evaluate precisely. However, because the trigger rates of both instruments are very different and the source flux was low (i.e. large effect of the random background on the resulting spectra), we expect the correction to be low and therefore treated the two experiments as independent.

\section{Results}
\label{sect:results}
We report the results of the \mbox{LST-1} data analysis in Sect.~\ref{sect:LST-1results} and compare them with MAGIC and H.E.S.S. findings. Additionally, the results of the model parameter fitting using Fermi-LAT, LST-1, MAGIC, and H.E.S.S. data are presented in Sect.~\ref{sect:modelling_results}, where we also compare the outcomes of the different models.

\subsection{\mbox{LST-1} results}
\label{sect:LST-1results}

The source RS~Oph was detected with \mbox{LST-1} with a statistical significance of 6.6$\sigma$ for the three days of LST-1 observations within the first four days from the nova eruption ($t-t_0\sim$ 1\,d, 2\,d, and 4\,d; see Table~\ref{tab:obs_table}). The daily statistical significance of the detection is discussed in Appendix~\ref{sect:sig} and shown in Table~\ref{tab:LST1_best_fit_model}. The source was not detected (1.6$\sigma$) using the observations conducted three weeks after the nova onset ($t-t_0>21\, \rm{d}$).

\begin{figure}[h!]
\centering
    \includegraphics[width=1\hsize]{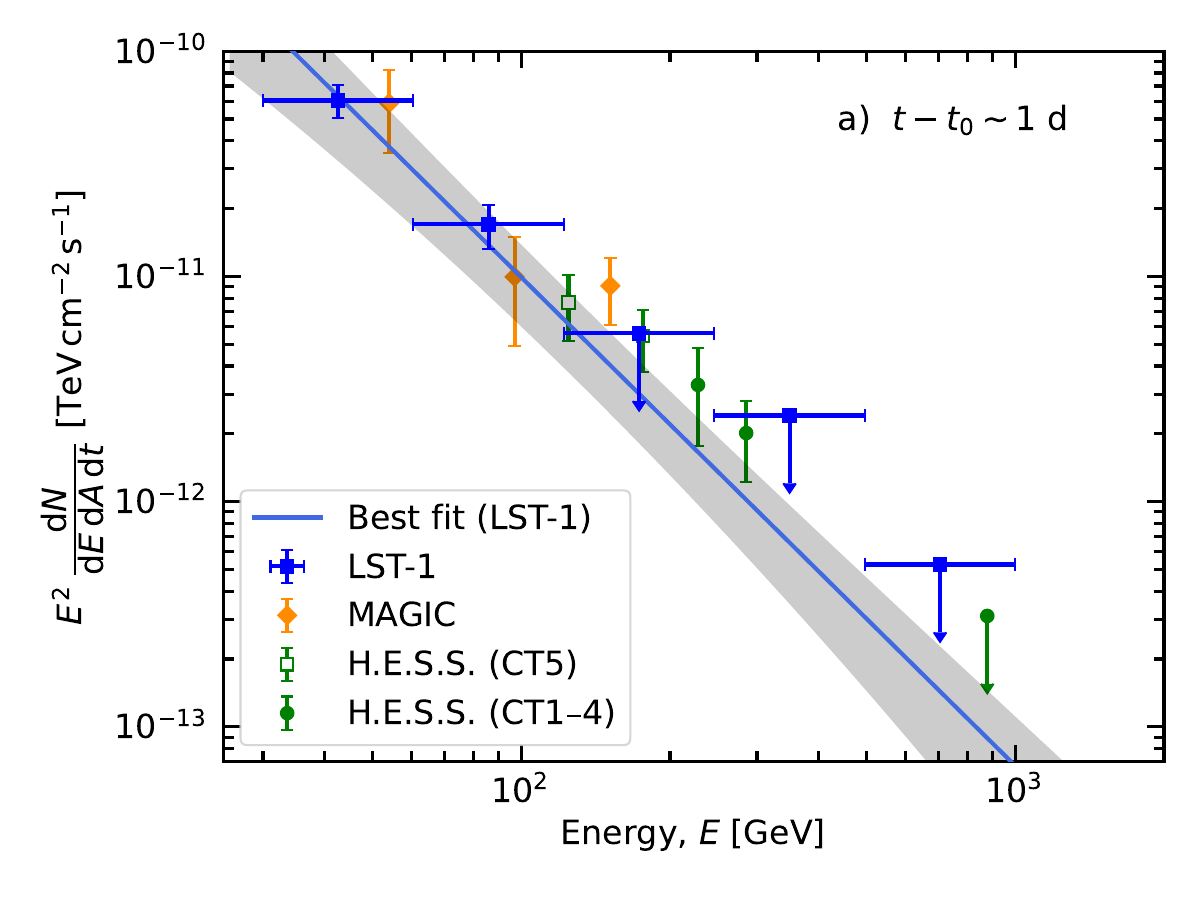}
    
    \includegraphics[width=1\hsize]{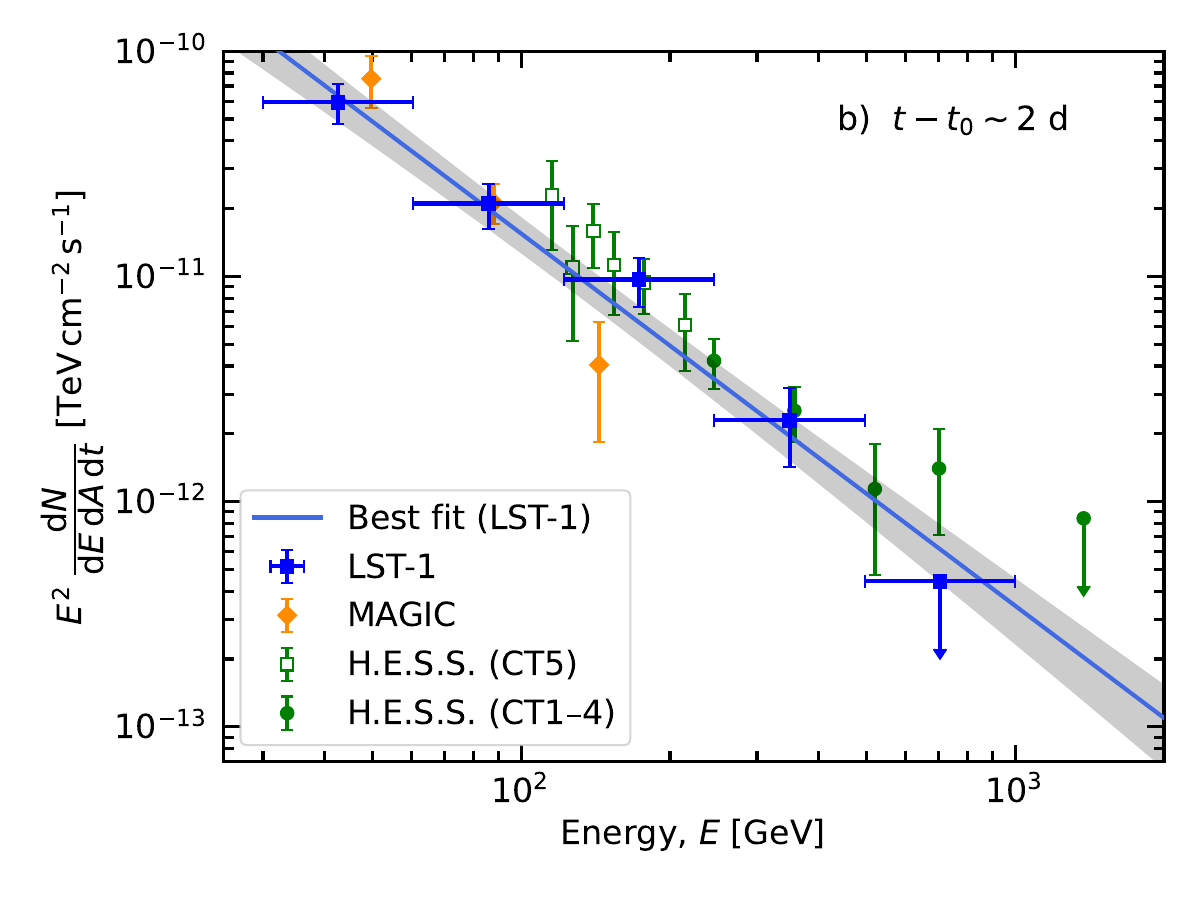}
    
    \includegraphics[width=1\hsize]{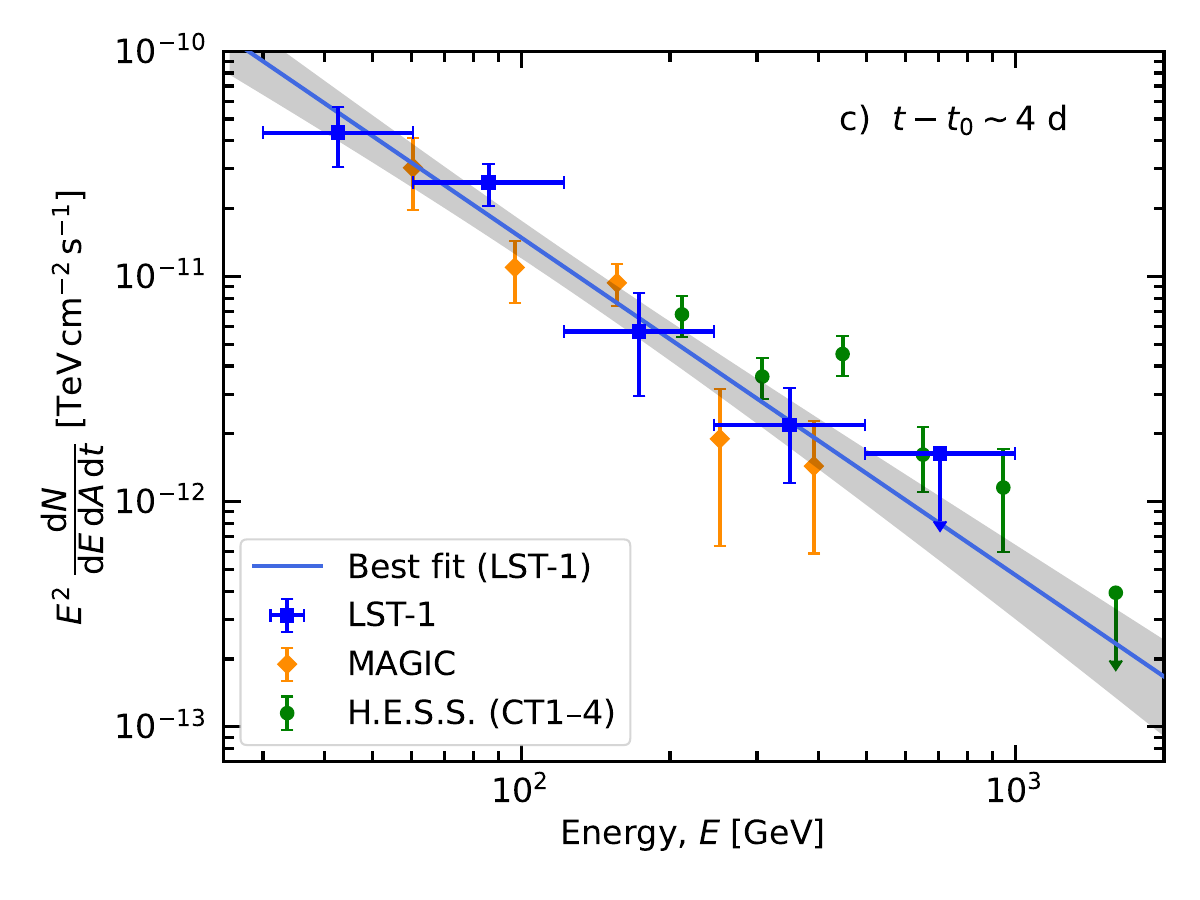}
    
\caption{RS~Oph daily SEDs at VHE gamma rays with \mbox{LST-1} (blue squares), MAGIC \citep[orange diamonds;][]{2022_RSOphMAGIC}, and \mbox{H.E.S.S.} \citep[green empty squares and filled circles for the telescopes CT5 and CT1--4, respectively;][]{2022_RSOphHESS} during the same day interval. From top to bottom, panels a, b, and c correspond to observation-day intervals $t-t_0 \sim$ 1\,d, 2\,d, and 4\,d, respectively. The best-fit model for \mbox{LST-1} is displayed as a blue line together with the grey spectral error band.
}
\label{fig:daily_SED_VHEs}
\end{figure}

For the purpose of aggregating data from different instruments, we defined observation days as the integer sequence of day intervals centred on $t_0$. The \textit{i}th observation-day interval spans over MJD $t_0+i\,\pm 12\,\mathrm{h}$. The RS~Oph daily SEDs at VHE gamma-ray energies with \mbox{LST-1}, including the best-fit spectral models, for the first three observations with LST-1 data on days 1, 2, and 4 after the explosion are shown as blue squares in Fig.~\ref{fig:daily_SED_VHEs}. A power-law spectral model was adopted to fit the \mbox{LST-1} data of RS~Oph. The spectral index measured with \mbox{LST-1} is soft for all days and seems to harden as the eruption evolves in time from observation day 1 to observation days 2--4 (see Table~\ref{tab:LST1_best_fit_model}). However, a spectral profile with a constant index set to the weighted average of the first four observation days cannot be rejected ($\textrm{p-value}=0.11$; a significance level\footnote{The maximum acceptable probability of committing a type I error assigned in the test statistic.} of $\alpha=0.05$ is set to test the null hypothesis). The situation is similar for a constant-amplitude model ($\textrm{p-value}=0.22$). 

\begin{table}
    \caption{
    Best-fit power-law spectral parameter values and statistical detection significances (Sig.) using \mbox{LST-1} data.
    }
    \begin{center}
        \footnotesize
        \label{tab:LST1_best_fit_model}
        \begin{tabular}{lr@{$~\pm~$}l@{~~}r@{$~\pm~$}lc}
        \hline \hline
        \noalign{\smallskip}
        Obs. day & \multicolumn{2}{c}{$\Gamma$}  & \multicolumn{2}{c}{$\phi_0$} & Sig. \\
         & \multicolumn{2}{c}{}  &  \multicolumn{2}{c}{[$10^{-10}\, \rm{TeV}^{-1}\, \rm{cm}^{-2}\, \rm{s}^{-1}$]} & [$\sigma$] \\

        \noalign{\smallskip}
        \hline
        \noalign{\smallskip}         
         Day 1 & $-4.2$~~ & $0.3$~~~~~~~ &  ~~~~~~~~~~~~$3.3$ & $1.3$ &  3.2\\
         Day 2 & $-3.65$ & $0.13$~ &  $5.9 $ & $ 1.0$ & 2.8\\
         Day 4 & $-3.50$ & $0.15$~ &  $5.9 $ & $ 1.1$ & 5.5 \\
         \noalign{\smallskip}
        \hline
        \noalign{\smallskip}
         Day 1, 2 and 4 & $-3.73 $ & $ 0.10$ &  $5.2 $ & $ 0.7$ & 6.6\\
        \noalign{\smallskip}         
         \hline
        
        \end{tabular}
    \end{center}
    \tablefoot{$\Gamma$ is the spectral index, and $\phi_0$ is the amplitude at a reference energy of $130\,\mathrm{GeV}$. The statistical detection significance \citep[using formula 17;][]{1983ApJ...272..317L} is computed for the full energy range (see Appendix~\ref{sect:sig} for more details).}    
\end{table}

In Fig.~\ref{fig:daily_SED_VHEs} we compare the SEDs obtained with \mbox{LST-1}, MAGIC, and \mbox{H.E.S.S.} during the same observation-day intervals \citep{2022_RSOphMAGIC, 2022_RSOphHESS}. 
In general, it is sufficient to only consider statistical errors to obtain compatible results between \mbox{LST-1} and \mbox{H.E.S.S.}/MAGIC, while \mbox{LST-1} spectra can probe lower energies than the other two IACTs. In addition to the daily SEDs shown in Fig.~\ref{fig:daily_SED_VHEs}, the joint spectrum for the same observation days ($t-t_0 \sim$ 1\,d, 2\,d, and 4\,d) with \mbox{LST-1} is provided in Appendix~\ref{sect:joint_analysis}. The joint \mbox{LST-1} SED is compatible with that of MAGIC during the same time period.

\begin{figure}
\centering
\includegraphics[width=1\hsize]{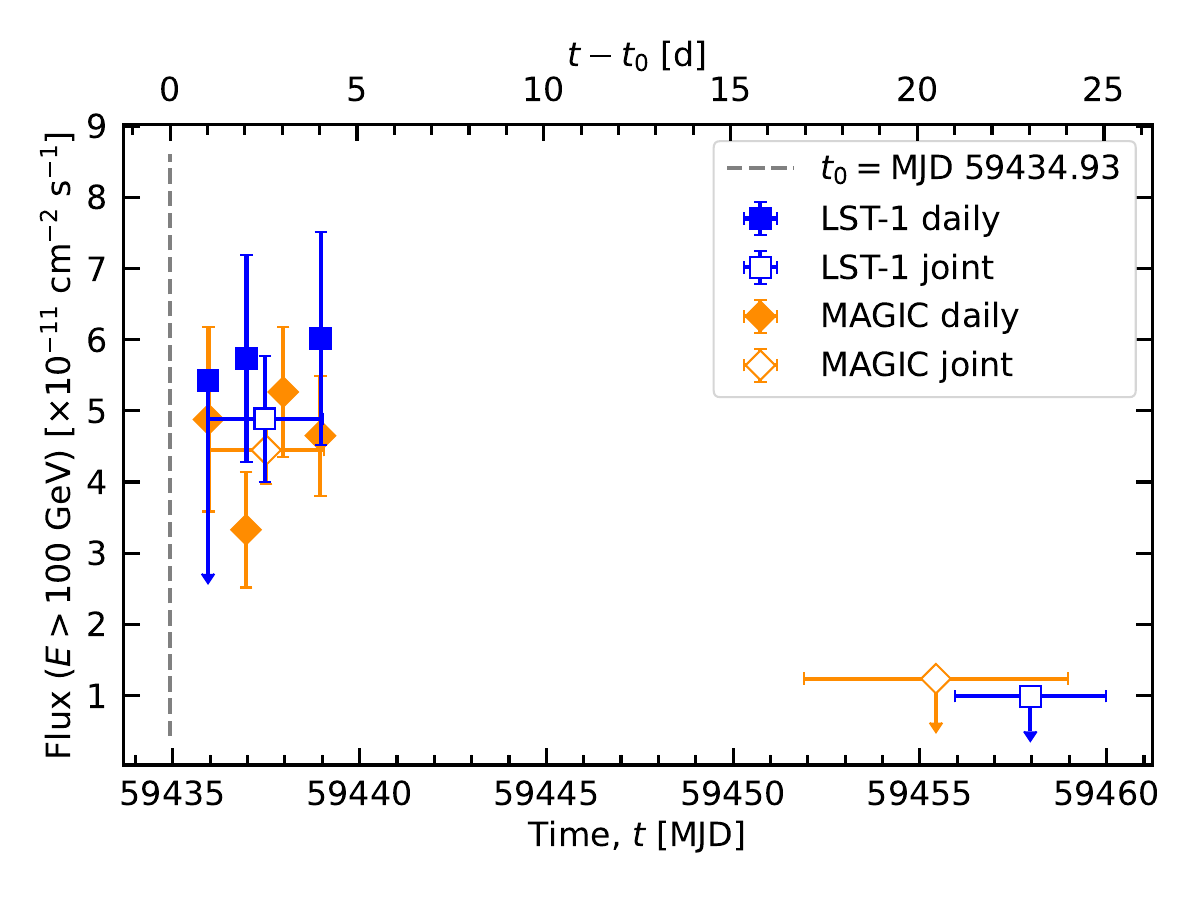}
  \caption{Daily integral fluxes at $E>100\, \text{GeV}$ for \mbox{LST-1} (filled blue squares) and MAGIC \citep[filled orange diamonds;][]{2022_RSOphMAGIC} as a function of MJD (bottom axis) and time after the eruption onset (top axis), which is represented as the dashed line. We also show the joint \mbox{LST-1} (empty blue squares) and the joint MAGIC (empty orange diamonds) integral fluxes during observation-day intervals 1 and 4, and more than 21 days after the eruption.
  }
     \label{fig:LC_LST-MAGIC}
\end{figure}

We show in Fig.~\ref{fig:LC_LST-MAGIC} the light curve above $100\,\mathrm{GeV}$ for \mbox{LST-1} and MAGIC \citep{2022_RSOphMAGIC}. We show the daily and joint integral fluxes for both instruments. No significant emission is detected with \mbox{LST-1} above 100\,GeV during the first day of data-taking (UL with $ TS=2.2$). The source flux $TS$ is instead above the UL criterion in the following days, with the LST-1 fluxes values compatible within the statistical uncertainties among them. We obtain less constraining spectral parameter values and a non-significant integral flux for the first observation day than for the second and fourth day because the observation time was shorter and the intrinsic emission was likely softer. The \mbox{LST-1} daily light curve differs slightly from that of MAGIC when we only consider statistical uncertainties, which suggests the presence of systematics that are not accounted for in the integral flux computation. When the data are joined, however, the joint VHE flux between observation-day intervals 1 and 4 with \mbox{LST-1} is compatible within the statistical uncertainties with the MAGIC joint integral flux for the same time window. We note that the MAGIC joint flux includes the observations on August 11 ($t-t_0 \sim 3\, \rm{d}$), when \mbox{LST-1} did not observe RS~Oph. For observations taken at $t-t_0>21\, \rm{d}$, when the source was not detected with \mbox{LST-1}, we computed a joint integral flux UL above $100\,\mathrm{GeV}$ of $10^{-11}\, \rm{cm^{-2}\,s^{-1}}$, which was obtained by fixing the power-law spectral index to $\Gamma=-3.5$ (the spectral index of observation day 4). The result is comparable in flux level with the MAGIC findings.

\subsection{Modelling results using \textit{Fermi}-LAT and IACT data}
\label{sect:modelling_results}

\begin{figure}[h!]
        \centering

        \includegraphics[clip,width=0.95\columnwidth]{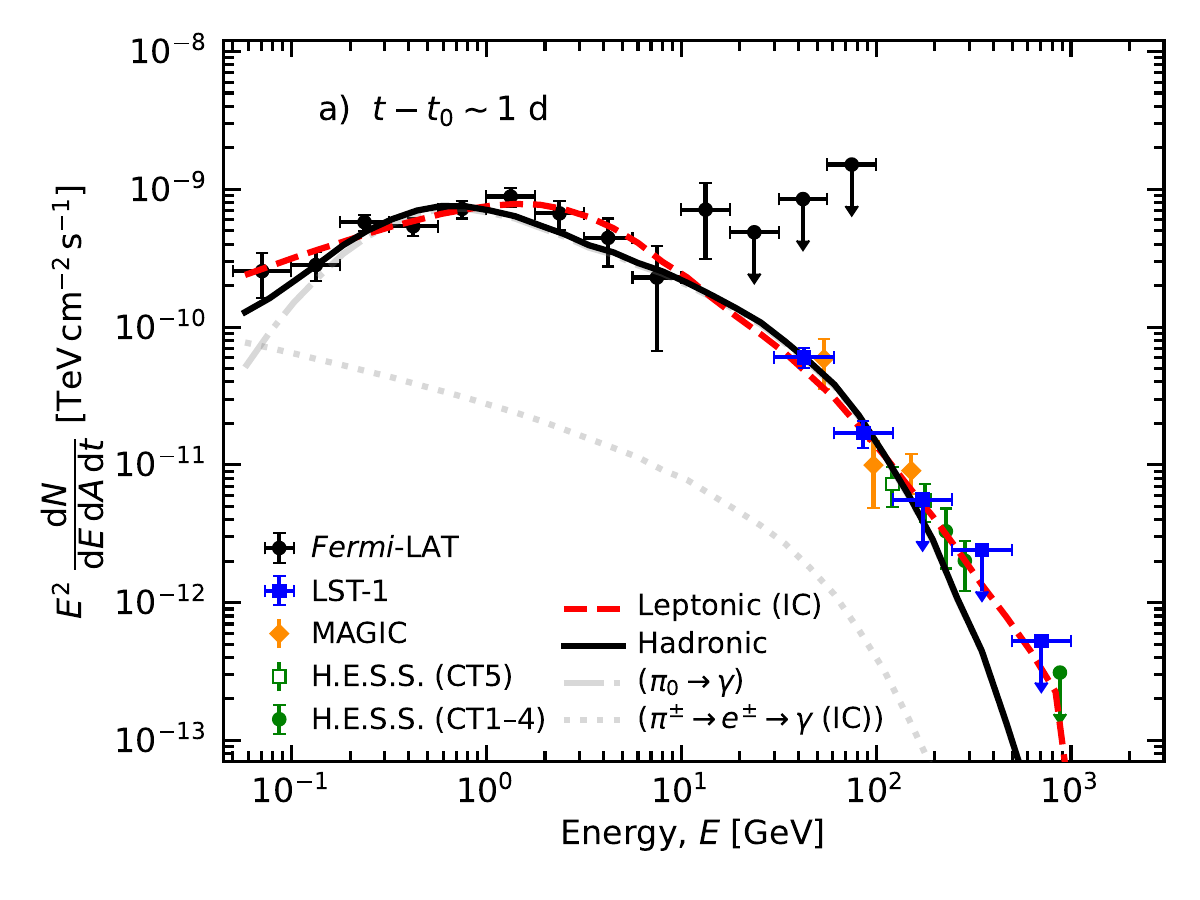}
        
        \includegraphics[clip,width=0.95\columnwidth]{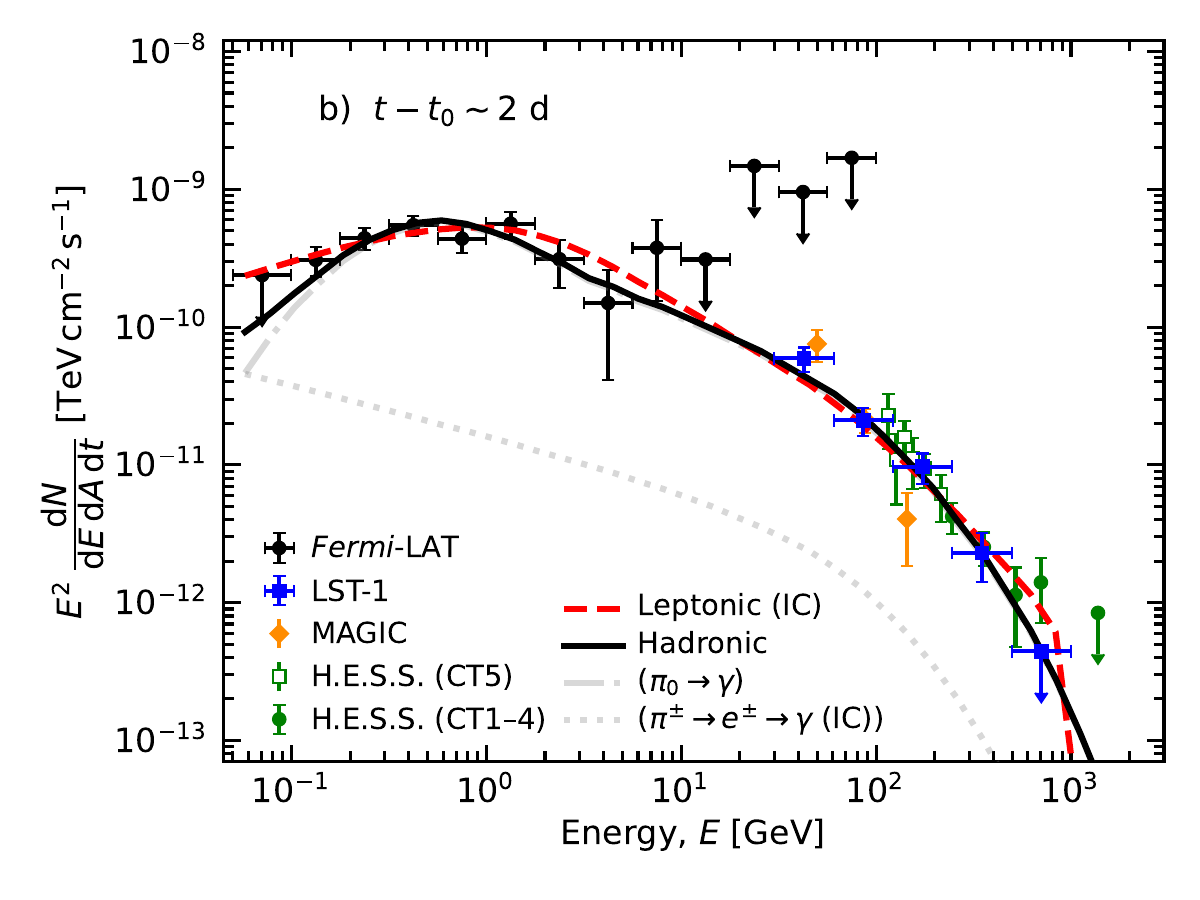}
        
        \includegraphics[clip,width=0.95\columnwidth]{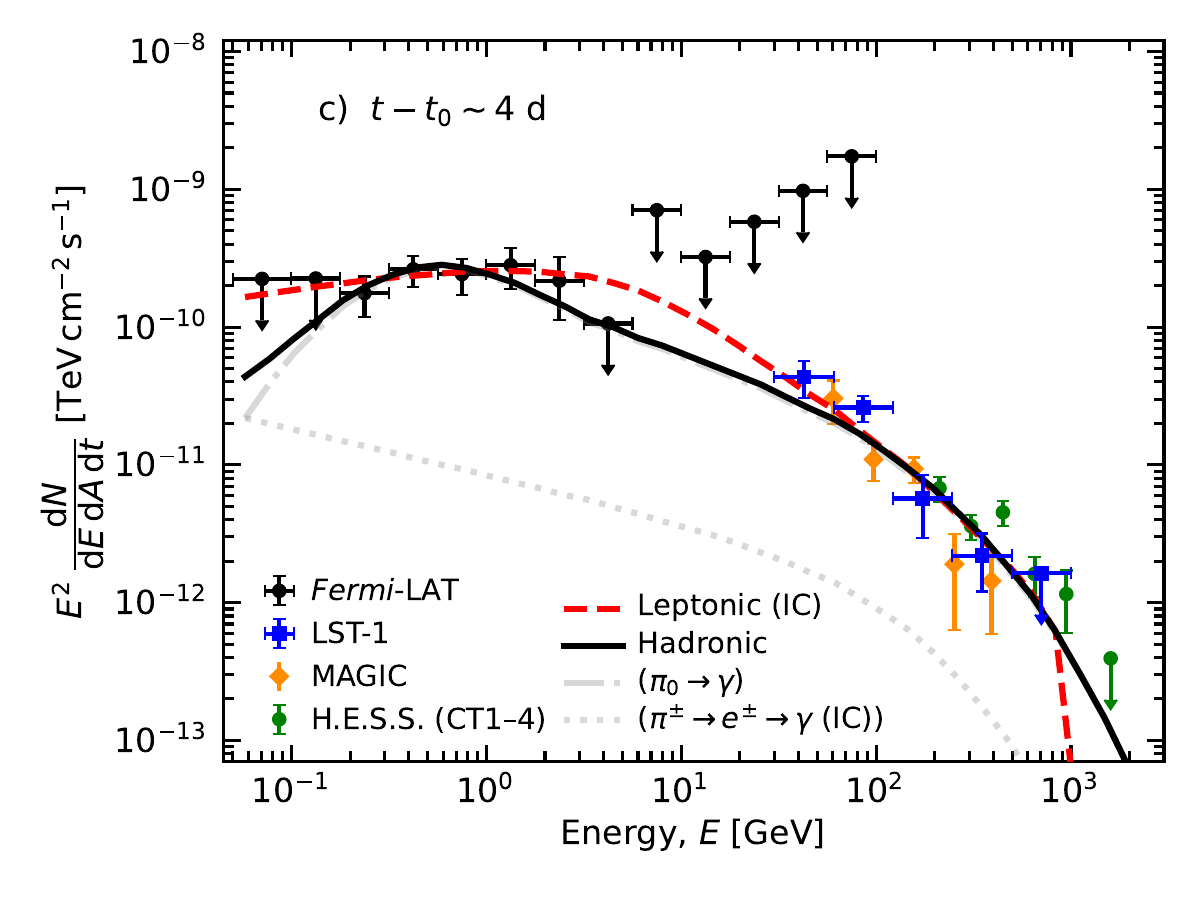}

        \caption{RS~Oph daily SEDs from HE to VHE gamma rays using \textit{Fermi}-LAT (black circles), \mbox{LST-1} (blue squares), MAGIC \citep[orange diamonds;][]{2022_RSOphMAGIC}, and \mbox{H.E.S.S.} \citep[green empty squares and filled circles for the telescopes CT5 and CT1--4, respectively;][]{2022_RSOphHESS}. From top to bottom, panels a, b and c correspond to observation-day intervals $t-t_0 \sim$ 1\,d, 2\,d, and 4\,d, respectively. The best-fit leptonic and hadronic models are displayed as dashed red and solid black curves, respectively. For the hadronic model, the corresponding contributions from neutral and charged-pion decays are shown in grey.
        }
            \label{fig:daily_modelling_FermiLST-MAGIC-HESS}        
\end{figure}

We show in Fig.~\ref{fig:daily_modelling_FermiLST-MAGIC-HESS} the daily SEDs with \textit{Fermi}-LAT together with the VHE IACT data that were shown in Fig.~\ref{fig:daily_SED_VHEs}. The gamma-ray spectra span from $50\,\mathrm{MeV}$ up to about $1\,\mathrm{TeV}$. The spectral information at HE connects smoothly with that of VHE gamma rays, showing signs of curvature.

The emission model was fitted using spectral information from \mbox{LST-1} and \textit{Fermi}-LAT data (this work) together with published MAGIC and \mbox{H.E.S.S.} spectral information for each coincident observation-day interval with \mbox{LST-1}. The simplest physically motivated particle energy profile, an ECPL model, without systematic energy-scale uncertainties in the fitting process (see Sect.~\ref{sect:modelling}), was considered for the hadronic and leptonic models. The ECPL model is not able to provide a good fit to the data for the leptonic model with reduced $\chi^2$ values for each day of about 3 ($\textrm{p-values}\sim 10^{-5}$). The poor fit can be explained by the spectral curvature present across the HE and VHE range (as already mentioned by \citealt{2022_RSOphMAGIC}).

To reproduce the curvature of the gamma-ray spectrum for the leptonic model, we used a BPL shape to describe the energy distribution of electrons, as suggested by \citet{2022_RSOphMAGIC}. The fluxes of the best BPL leptonic models are shown as red curves on the daily SEDs in Fig.~\ref{fig:daily_modelling_FermiLST-MAGIC-HESS}, while the corresponding fit results are displayed in Table~\ref{tab:model_fitting_results}. The leptonic model with a BPL spectral shape can describe the curvature of the spectrum with \textit{Fermi}-LAT and IACT data. As the SED evolves with time from observation day 1 to 4, the best-fit parameters of the leptonic model evolve as well: The slope below the energy break softens while the energy break shifts towards higher energies; the slope above the energy break hardens with time. Overall, the electron spectrum reaches higher energies with time. However, the uncertainties of the best-fit parameters of the leptonic model shown in Table~\ref{tab:model_fitting_results} are large because the fitting process is rather dependent on the input parameter values (more information on the robustness of the fitting can be found in Appendix~\ref{sect:model_fit_results}). The spectral parameter results are compatible within the uncertainties with the best-fit leptonic model by \citet{2022_RSOphMAGIC}.

We now consider the ECPL hadronic model to fit the data. The emission of the best model is shown as black curves on the daily SEDs in Fig.~\ref{fig:daily_modelling_FermiLST-MAGIC-HESS}, while the corresponding fit results are displayed in Table~\ref{tab:model_fitting_results}. The spectral index softens as the eruption evolves, while the proton spectrum cutoff energy increases with time: from $(0.26\pm 0.08)\,\mathrm{TeV}$ on day 1 to $(1.6 \pm 0.6)\,\mathrm{TeV}$ on day 4. A constant cutoff energy during the first three LST-1 observation days is rejected with a $\textrm{p-value}=6\times10^{-3}$. The temporal evolution of the proton shape is similar but not compatible within the uncertainties to the one observed with \textit{Fermi}-LAT and MAGIC alone \citep{2022_RSOphMAGIC}. The hadronic fit of all gamma-ray data presents softer proton spectra and higher cutoff energies.

\begin{table}[h!]
    \caption{Model-fit results without and with systematic uncertainties using the leptonic and hadronic modelling for observation days $t-t_0 \sim1$\,d, 2\,d, and 4\,d.
    }
    \begin{center}
        \footnotesize
        \label{tab:model_fitting_results}
        {\color{black}\begin{tabular}{lccc}
        \hline
        \hline
        \noalign{\smallskip}
        Parameter  & \multicolumn{3}{c}{Best-fit value on observation day} \\
        \cline{2-4}
                \noalign{\smallskip}

                   &  Day 1  &  Day 2  &  Day 4 \\
        \noalign{\smallskip}
        \hline
        \noalign{\smallskip}

        \multicolumn{4}{c}{Leptonic BPL model without systematics} \\
        \noalign{\smallskip}
        \hline
        \noalign{\smallskip}
        Slope 1, $\Gamma_{\rm e,1}$ & $0.0^{+0.8}_{-0.6}$~ & $-1.3^{+1.5}_{-0.5}$~~~ & $-1.4^{+0.6}_{-0.7}$~~~ \\
        \noalign{\smallskip}
        Slope 2, $\Gamma_{\rm e,2}$ & $-3.79^{+0.17}_{-0.18}$ &  $-3.57^{+0.11}_{-0.15}$ & $-3.52^{+0.05}_{-0.06}$ \\
        \noalign{\smallskip}
        $E_{\rm b,e}$ [$\rm{GeV}$] & $14^{+3}_{-3}$ &  $17^{+8}_{-4}$ & $22^{+9}_{-6}$ \\    
        \noalign{\smallskip}        
        $\chi^2/N_{\rm d.o.f}$ & $12.9/15$ & $24.9/21$ & $24.9/15$ \\
        \noalign{\smallskip}        
        $\chi^2_{\rm red}$ & $~~~~0.86$ & $~~~~1.19$ & $~~~~1.66$ \\        
        \noalign{\smallskip}
        AIC$_{\rm c}$ & $23.7$ & $34.9$ & $35.8$ \\
        \noalign{\smallskip}                
        \hline        
        \noalign{\smallskip}

        \multicolumn{4}{c}{Hadronic ECPL model without systematics} \\        
        \noalign{\smallskip}
        \hline
        \noalign{\smallskip}      
        Slope, $\Gamma_{\rm p}$ & ~$-2.25^{+0.13}_{-0.13}$ & $-2.49^{+0.07}_{-0.07}$ & $-2.48^{+0.08}_{-0.08}$ \\
        \noalign{\smallskip}
        $E_{\rm c,p}$ [$\rm{TeV}$] &  ~~~$0.26^{+0.08}_{-0.08}$ & $1.0^{+0.3}_{-0.3}$~ & $1.6^{+0.6}_{-0.6}$~~\\
        \noalign{\smallskip}
        $\chi^2$/$N_{\rm d.o.f}$ & $21.5/16$ & $24.9/22$ & $26.5/16$ \\
        \noalign{\smallskip}
        $\chi^2_{\rm red}$ & ~~~~$1.34$ & ~~~~$1.13$ & ~~~~$1.66$ \\     
        \noalign{\smallskip}
        AIC$_{\rm c}$ & $29.1$ & $32.0$ & $34.1$ \\        
        \noalign{\smallskip}        
        \hline
        \noalign{\smallskip}

        \multicolumn{4}{c}{Leptonic BPL model with systematics} \\
        \noalign{\smallskip}
        \hline
        \noalign{\smallskip}
        Slope 1, $\Gamma_{\rm e,1}$ & $~~\,0.4^{+1.9}_{-1.9}$~~~ & $-1.6^{+0.8}_{-0.3}$ & $-1.4^{+0.8}_{-0.7}$~~~ \\
        \noalign{\smallskip}
        Slope 2, $\Gamma_{\rm e,2}$ & $-3.70^{+0.17}_{-0.17}$ &  $-3.6^{+0.2}_{-0.2}$ &  $-3.75^{+0.13}_{-0.11}$\\
        \noalign{\smallskip}
        $E_{\rm b,e}$ [$\rm{GeV}$] & $13^{+3}_{-3}$ &  $20^{+9}_{-8}$ &  $30^{+11}_{-10}$ \\    
        \noalign{\smallskip}
        $\chi^2/N_{\rm d.o.f}$ & $12.9/12$ & $22.8/18 $ & $16.8/12$\\
        \noalign{\smallskip}
        $\chi^2_{\rm red}$ & ~~~~$1.08$ & ~~~~$1.27$ & ~~~~$1.40$ \\       
        \noalign{\smallskip}
        AIC$_{\rm c}$ & $37.1$ & $43.4$ & $41.0$ \\   
        \noalign{\smallskip}        
        \hline
        \noalign{\smallskip}  

        \multicolumn{4}{c}{Hadronic ECPL model with systematics} \\
        \noalign{\smallskip}
        \hline
        \noalign{\smallskip}
        Slope, $\Gamma_{\rm p}$ & $-2.22^{+0.06}_{-0.10}$ &  $-2.51^{+0.05}_{-0.05}$ & $-2.40^{+0.15}_{-0.15}$ \\
        \noalign{\smallskip}
        $E_{\rm c,p}$ [$\rm{TeV}$] & ~~\,$0.23^{+0.06}_{-0.04}$ &  $0.9^{+0.2}_{-0.2}$\, & $1.0^{+0.6}_{-0.6}$\, \\
        \noalign{\smallskip}        
        $\chi^2/N_{\rm d.o.f}$ & $21.1/13$ & $20.4/19$ & $19.9/13$ \\
        \noalign{\smallskip}    
        $\chi^2_{\rm red}$ & ~~~~$1.62$ & ~~~~$1.07$ & ~~~~$1.53$ \\
        \noalign{\smallskip}
        AIC$_{\rm c}$ & $40.1$ & $37.1$ & $38.9$ \\ 
        \noalign{\smallskip}        
        \hline              
        \end{tabular}}
    \end{center}
    \tablefoot{
    For the leptonic modelling, $\Gamma_{\rm e,1}$ and $\Gamma_{\rm e,2}$ are the best-fit slopes below and above the best-fit energy break ($E_{\rm b,e}$), respectively, of the electron energy distribution. For the hadronic case, $\Gamma_{\rm p}$ is the best-fit slope and $E_{\rm c,p}$ is the best-fit cutoff energy of the proton energy distribution. We provide the $\chi^2_{\rm red}$ fit statistics ($\chi^2_{\rm red}=\chi^2/N_{\rm d.o.f}$) and the daily AIC$_{\rm c}$ values (see text). The sum of the AIC$_{\rm c}$ values for all days for the leptonic model without and with systematics is 94.4 and 121.5, respectively, while for the hadronic model without and with systematics, it is 95.2 and 116.1, respectively. The error values correspond to the quadratic sum of 1$\sigma$ fit and sampling errors (Appendix~\ref{sect:model_fit_results}). The units of $E_{\rm c,p}$ and $E_{\rm b,e}$ are in TeV and GeV, respectively.
    }
\end{table}

The Akaike information criterion (AIC) estimator \citep{Akaike_1974} was used to compare different non-nested models. We summed the AIC values corrected for the second-order small-sample bias adjustment \citep[AIC$_{\rm c}$;][]{hurvich1989regression} of all observation days for the hadronic and leptonic models ($\sum$AIC$_{\rm c,p}$ and $\sum$AIC$_{\rm c,e}$, respectively) and computed the difference of the higher to the lowest one ($\sum$AIC$_{\rm c,min}$) to compare them ($\Delta\rm{AIC_{c}}$). There is no preference between the hadronic over the leptonic model: $\Delta\rm{AIC_{c}}=0.8$, which corresponds to a relative likelihood \citep{AKAIKE19813} of $0.7$, where $\sum$AIC$_{\rm c,min}=\sum$AIC$_{\rm c,e}$. 
The slight loss of information, that is, $\Delta\rm{AIC_{c}}>0$, experienced by using the hadronic over the leptonic model comes from the low $\chi^2$ value for the leptonic model on observation day 1. However, on this day, the best-fit leptonic model presents the strongest spectral slope break ($\Gamma_{\rm e,1}-\Gamma_{\rm e,2} = 3.8^{+1.0}_{-0.8}$, where $\Gamma_{\rm e,1}$ and $\Gamma_{\rm e,2}$ are the slope below and above the energy break, respectively). The no-preference of the hadronic model based on the fit statistics disagrees with the results shown in \citet{2022_RSOphMAGIC}.

We note that the $\chi^2_{\rm red}$ for all models deviates from one. Remarkable spectral discrepancies at hundreds of GeV are noticeable between the MAGIC and \mbox{H.E.S.S.} differential fluxes, the latter presenting a harder gamma-ray emission than MAGIC (e.g. see Fig.~\ref{fig:daily_SED_VHEs}). The mismatch between MAGIC and \mbox{H.E.S.S.} results likely contributes to worsening the goodness of fit of the models. Thus, to account for the \mbox{LST-1}, MAGIC, and \mbox{H.E.S.S.} discrepancies due to possible energy-scale uncertainties in the IACT data analyses, the hadronic and leptonic models were refitted including systematic energy-scale uncertainties as nuisance parameters in the fitting process (see Appendix~\ref{sect:fitting_with_syst}). The best-fit results are shown in Table~\ref{tab:model_fitting_results}. The hadronic and leptonic results with and without systematic are compatible and exhibit the same temporal trends of the particle spectra.

We summed the AIC$_{\rm c}$ values of all the observation days and compared the models with and without systematics. The leptonic and hadronic model fits without systematics are both favoured over considering them ($\Delta\mathrm{AIC_c}=27.1$ and $ \Delta\mathrm{AIC_c}=20.9$, with a relative likelihood of $ 1\times 10^{-6}$ and $ 3\times 10^{-5}$ for the leptonic and hadronic models, respectively). Therefore, we discarded the fit results accounting for energy-scale systematics during the model fitting and consider hereafter the leptonic and hadronic model without systematics to describe the relativistic particle energy distribution of RS~Oph.

More complicated models can be considered to explain the RS Oph gamma-ray emission. A population of both electrons and protons was used with the lepto-hadronic model in \citet{2022_RSOphMAGIC}, even though it was not preferred due to a poor fit and an order of magnitude larger proton-to-electron luminosity ratio (Lp/Le) than the constrained one in the classical nova V339~Del \citep{2022_RSOphMAGIC}. We note that although HE gamma rays are produced in RS~Oph and V339~Del, the constrained Lp/Le in the latter may not apply to RS Oph because classical novae such as V339~Del may have different HE particle distribution than embedded novae because the particle acceleration mechanisms and shock formation regions might differ. 

The details and results of the lepto-hadronic model are provided in Appendix~\ref{sect:lephad}. Based on the fit statistics, the lepto-hadronic model is substantially less supported and presents similar parameter values as the hadronic results obtained for observation days $t-t_0 \sim1\,\textrm{d}$ and 4\,d, while for the second observation day, the leptonic component dominates the HE band. Therefore, we considered that the lepto-hadronic model is less plausible, based on gamma-ray data alone, to explain the gamma-ray spectrum.

\section{Discussion and outlook}
\label{sect:discussion_outlook}

The results above show that RS~Oph was detected with \mbox{LST-1} during several coincident days with \textit{Fermi}-LAT, MAGIC, and \mbox{H.E.S.S.} In this section, we discuss the spectral and modelling results of RS~Oph, and we provide an outlook of future expectations for nova detections with the array of LSTs of CTAO.

\subsection{Discussion}
\label{sect:discussion}

The data from \textit{Fermi}-LAT, LST-1, MAGIC, and H.E.S.S. allowed us to study the emission from $\sim$50\,MeV up to multi-TeV energies. This is the largest HE-VHE dataset compiled to date, for which a total exposure time of 6.35\,h, 7.7\,h, and 9.2\,h (6.0\,h) is considered for LST-1, MAGIC, and H.E.S.S. CT1--4 (CT5), respectively. While H.E.S.S. constrains the gamma-ray spectrum at about 1 TeV because its sensitivity is better at these energies than that of MAGIC and LST-1, LST-1 bridges the HE and VHE gamma-ray emission and reducing the lower energy bound of the SED of IACTs to $\sim$30 GeV. This is lower by a factor of $\sim$2 than the MAGIC energy threshold \citep{2022_RSOphMAGIC}. However, the sensitivity of LST-1 is worse by a factor of $\sim$1.5 than that of MAGIC above 100\,GeV because of the advantages of the stereoscopic reconstruction mode in MAGIC \citep{LSTPerformance}.

The model fitting with the \textit{Fermi}-LAT, \mbox{LST-1}, MAGIC, and \mbox{H.E.S.S.} spectral information describes the spectrum of RS~Oph during the first days after the eruption onset using a one-zone single-shock model. 
Although the hadronic model was slightly preferred statistically over the leptonic model by a combined \textit{Fermi}-LAT and single IACT data fitting \citep{2022_RSOphMAGIC}, there is no clear preference, based on the fit statistics, between the hadronic and leptonic model using the spectral information from all instruments together.

The simplest ECPL model for the leptonic model cannot adequately describe the curvature of the gamma-ray spectral shape. We also considered the BPL energy distribution shape in the leptonic model to account for this curvature. The physical meaning of using a BPL model is to account for two populations of relativistic electrons as the particles that cause the gamma-ray emission. Therefore, the curvature of the spectrum, which a single population of electrons has difficulties reproducing through IC cooling, is caused by the two populations of relativistic electrons.
The leptonic scenario with a BPL model presents several difficulties with respect to the ECPL hadronic model, however. Firstly, our leptonic modelling presents multiple local minima that influence the fitting procedure through the relatively large number of free model parameters. The distribution of the best-fit first slope values ($\Gamma_{\rm e,1}$) for observation day 1 is bimodal with two peaks, which complicates the estimation of the confidence interval of this parameter and the $\chi^2$ statistics (see Appendix~\ref{sect:model_fit_results} for details). Secondly, the best-fit electron distributions have a difficult physical interpretation: They are characterised by a very different spectral slope below and above the energy break, as pointed out by \citet{2022_RSOphMAGIC}. In particular, the slope break is more pronounced for $t-t_0 \sim1\,\textrm{d}$, when the emission at HE gamma-rays is bright and the curvature of the spectrum is significantly visible: A flat HE emission is rejected with a p-value of $ 7\times10^{-5}$ using the \textit{Fermi}-LAT SED points presented in this work.
A very hard (positive or flat) $\Gamma_{\rm e,1}$ index of the electron energy distribution suggests that the fit on the first day tries to imitate the injection of electrons with a high minimum energy ($\sim14$~GeV). 
The injection of monoenergetic HE particles like these in novae was investigated (in the case of protons) by \citet{2022MNRAS.511.3339B}.
Nevertheless, on observation day 4, the spectrum below the break is compatible within the uncertainties to the classical $-2$ slope obtained in diffusive shock acceleration (DSA) for strong shocks \citep{1983RPPh...46..973D}.
There is no straightforward explanation for why on this day, a steep $\sim -3.5$ spectrum also needs to be injected above the break, and why the break energy is comparable (within uncertainties) to the injection energy from the first day of the nova. 
This picture of the required evolution of the electron energy distribution might be affected by partially cooled-down electrons that were injected at earlier phases.
However, as  showed by \citet{2022_RSOphMAGIC}, the GeV electrons in RS~Oph cool down on sub-day time scales.
We therefore conclude that while it is possible to describe the gamma-ray observations of RS~Oph with a leptonic model, the required injection electron energy distribution is not the one expected from DSA. This disfavours such a model. 
In consequence, we consider the hadronic scenario to be the most suitable mechanism for explaining the gamma-ray emission.

In the hadronic model, the average total power in protons across the first observation days ($t-t_0 \sim1$\,d, 2\,d and 4\,d) is $4.3\times 10^{43}\,\textrm{erg}$. For a kinetic energy of $2.0\times 10^{44}\,\textrm{erg}$ \citep{2022_RSOphMAGIC}, this indicates that the conversion efficiency of the shock energy to proton energies is about $20\%$. This considerable amount is aligned with the conversion efficiency estimated by \citet{2022_RSOphMAGIC} and above the lower limit estimated by \citet{2022_RSOphHESS}.

Systematic uncertainties in the energy scale were introduced in the model fitting to account for systematic errors in the IACT spectral results and reduce their impact on the estimated particle spectrum. However, the model with systematics is less favourable than the model without them, as the addition of three additional free parameters does not significantly improve the goodness of fit. Even though this does not imply that the dataset is free from systematic uncertainties, we chose to discard the model results that accounted for systematics as nuisance parameters. The consistent particle spectra between the hadronic models, both with and without systematics, suggest that the energy-scale uncertainties in the IACT data do not substantially affect the fit results. The primary limitations appear to originate from the small sample size and large flux uncertainties. Additionally, the limited information available from the public data points of MAGIC and H.E.S.S. complicates the application of more refined models that account for upper limits, which provides additional information. While alternative methods might address the systematics between IACTs, these are beyond the scope of this study. Although we discarded the results with systematics, it is worth noting that the sign and time evolution of the best-fit MAGIC and \mbox{H.E.S.S.} systematic energy scaling factors agree with the mismatch between the spectrum from MAGIC and \mbox{H.E.S.S.} and the fact that the gamma-ray emission at VHEs increases in time. This highlights the difference more strongly (see Appendix~\ref{sect:fitting_with_syst}).

We consider the best-fit proton spectrum and the subsequent gamma-ray emission from the hadronic modelling without systematics as the reference spectrum for RS~Oph. The best-fit proton spectra indicate that protons increase the maximum energy that can reach, up to TeV energies, in agreement with the estimated maximum proton energy and the temporal evolution in the same time period by \citet{2022ApJ...935...44C, 2022_RSOphMAGIC, 2022_RSOphHESS} via DSA. The evolution of the proton spectrum in time is interpreted as the finite acceleration time required for the protons in the expanding shock to accelerate from hundreds of GeV to TeV energies.

Other approaches were used to reproduce the spectral and temporal features observed in the RS~Oph eruption: a multi-population particle scenario of electrons and protons \citep{2022_RSOphMAGIC,De_Sarkar_2023}, or a multi-shock scenario approach \citep{Diesing_2023}. 

On the one hand, related to the multi-population particle scenario of electrons and protons, the non-thermal radio detection at early times supports the presence of relativistic electrons that rise early during the eruption for both classical and embedded novae \citep[e.g.,][]{2014Natur.514..339C, 2018ApJ...852..108F, 2016MNRAS.457..887W, 2006Natur.442..279O, 2024MNRAS.534.1227M}. In particular, non-thermal radio emission has been detected coincident with HE gamma rays in embedded ones \citep{2017ApJ...842...73L, 2023MNRAS.523..132D, 2023MNRAS.523.1661N}. For instance, synchrotron emission was detected coincident in time with the gamma-ray emission in RS~Oph \citep{2023MNRAS.523..132D} and the gamma-ray nova candidate V1535~Sco \citep{2017ApJ...842...73L,2018A&_Franckowiak}. However, the gamma-ray contribution from IC losses at early times is unclear. 
A detailed multi-frequency follow-up monitoring is required to obtain a precise spectrum to constrain the physics that cause the broad non-thermal emission. When the lepto-hadronic model in \citet{2022_RSOphMAGIC} was fitted with all available gamma-ray data, the HE component was described by IC, but with a hard spectral index, and it was disfavoured with respect to the leptonic and the hadronic models by the fit statistics. This may imply that there are further components within the very early gamma-ray emission from recurrent nova. This goes beyond the scope of this study, however, and requires early multi-wavelength observations of nova and additional modelling. 

On the other hand, connected to the multi-shock scenario, the non-spherical ejecta observed in RS~Oph and expected in embedded novae driven by the secondary star \citep{2022A&A...666L...6M, 2016MNRAS.457..822B, 2017MNRAS.464.5003O,2024ApJ...960..125I}, the multiple velocity components in the 2021 RS Oph nova \citep{Diesing_2023}, and a possible localised shock-acceleration event \citep{2022ApJ...935...44C} support the idea of a system with multiple shocks that evolve during the eruption. However, a single hadronic population model with spherical symmetry can reproduce the observed spectrum and temporal evolution at HE and VHE gamma-ray energies with acceptable accuracy (see Table~\ref{tab:model_fitting_results}). Furthermore, the temporal evolution of the proton energy distribution seems reasonably explained by the finite acceleration time for particles to reach TeV energies. To shed light on this matter, the observation of future bright novae with detectors with better sensitivities and deeper monitoring campaigns could help distinguish between the acceleration mechanisms and the contribution of non-thermal emission from a multiple particle population scenario (single or multiple shocks in a hadronic or lepton-hadronic scenario).

\subsection{Outlook}
\label{sect:outlook}

\begin{figure}
\centering
\includegraphics[width=1\hsize]{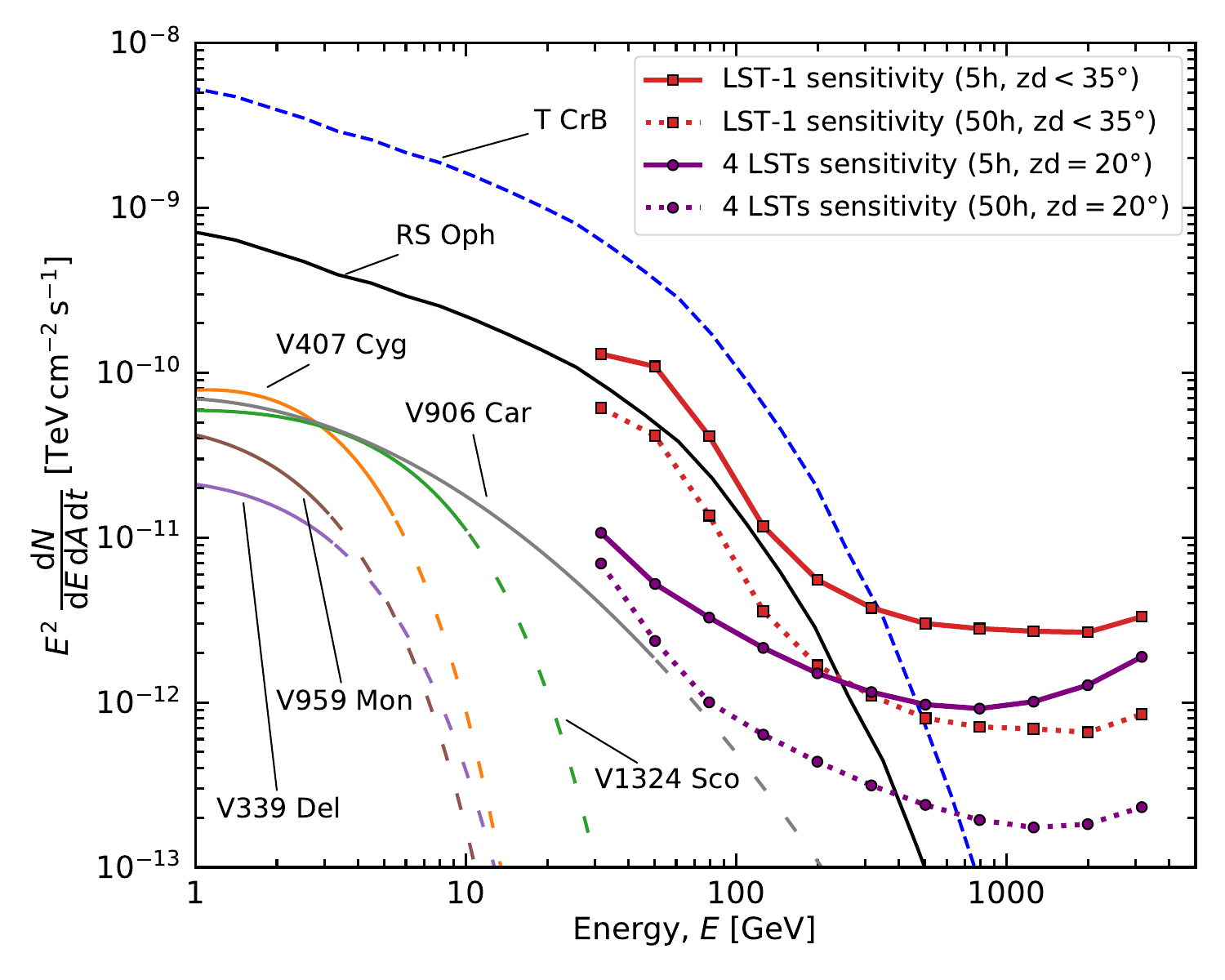}
  \caption{Best-fit SED models for RS~Oph (black; this work, $t-t_0 \sim 1\, \rm{d}$), V906~Car \citep[grey;][]{V960Car2020NatAs}, the first novae detected with \textit{Fermi}-LAT \citep[V407 Cyg, orange; V1324 Sco, green; V959 Mon, brown; V339 Del, pink;][]{FermiNovae2014Sci}, and the expected SED from T~CrB (blue). The solid region of the nova SEDs approximately corresponds to the energy range where the model was fitted to the data, while the spaced-dashed region corresponds to the extrapolation region of the model. In addition, the sensitivity curves for \mbox{LST-1} at low zenith angles \citep[$\rm zd<35$\degr; red square curves;][]{LSTPerformance} and the four LSTs of CTAO-North at $\rm zd=20$\degr\ (purple circle curves; using the latest \texttt{Prod5-v0.1} IRFs in \citealt{cherenkov_telescope_array_observatory_2021_5499840}) for 5 and 50 hours of observation time (see text for details) are displayed as solid and dotted curves, respectively.}
     \label{fig:sensitivity_LSTs}
\end{figure}

The source RS~Oph belongs to a specific nova class in which a binary system with a giant donor companion star undergoes recurrent outbursts. This class contains a few members: T~CrB, V3890~Sgr, V745~Sco, and RS~Oph \citep{2012clno.book.....B}. T~CrB is the closest system to Earth from this class \citep[$\sim 0.9\,\rm{kpc}$;][]{2023A&A...674A...1G}, followed by RS~Oph. T~CrB is expected to erupt again in the mid-2020s \citep{2023MNRAS.524.3146S,2023ATel16107....1S,2020ApJ...902L..14L}. When we assume that T~CrB will manifest the same spectral profile as RS~Oph, T~CrB will likely present a brighter flux than RS~Oph by a factor of $\sim 7$, making T~CrB one of the brightest novae at gamma rays up to date. In Fig.~\ref{fig:sensitivity_LSTs}, the best-fit SED models of seven embedded (V407~Cyg, T~CrB, and RS~Oph) and classical (V906~Car, V959~Mon, V1324~Sco, and V339~Del) novae detected at HE gamma rays are displayed together with the tentative gamma-ray SED from T~CrB based on the RS~Oph spectral shape on day 1, with the flux amplitude scaled to account for the difference in distance between them. The observed SED for the different novae is diverse, possibly due to their intrinsic nature and/or distance. Moreover, the observed SED models were estimated considering different observation time spans depending on the duration of the gamma-ray detection\footnote{V906 Car SED model does not satisfy this criterion because observations were restricted due to solar panel issues.}.  
Between 17 and 27 days of data were used to produce the gamma-ray spectral models in Fig.~\ref{fig:sensitivity_LSTs} \citep[except RS~Oph;][]{V960Car2020NatAs, FermiNovae2014Sci}. Hence, the displayed gamma-ray spectrum cannot accurately describe the maximum flux level or any spectral variability during the nova events. In addition, the gamma-ray flux for the \textit{Fermi}-LAT novae was extrapolated to the CTAO energy range assuming the best-fit spectral shape reported with \textit{Fermi}-LAT, consisting of an ECPL for the first novae detected with \textit{Fermi}-LAT. In contrast, a log-parabola model was considered for V906~Car, which resembles the RS~Oph spectral shape, but has a lower flux. The expected bright gamma-ray flux of T~CrB 
should help us to constrain the parameters of the particle population that causes the gamma-ray emission in novae even better.

It is still an open question whether classical novae can emit VHE gamma rays or if a bright VHE gamma-ray emission requires a similar binary system configuration as RS Oph, a recurrent nova embedded in the red giant star envelope. Related to the former, the brightest classical nova detected at HE gamma rays so far is V906~Car \citep{V960Car2020NatAs}. 
Its best-fit SED model is displayed in Fig.~\ref{fig:sensitivity_LSTs} together with the detection sensitivity curves for LST-1 and the array of four LSTs of CTAO-North for an integration time of 5\,h and 50\,h. The standard definition of sensitivity is used: The minimum flux for a 5$\sigma$ detection, with a requirement of a signal-to-background ratio of 5\% at least. This definition only applies to point-like sources calculated in a five-bins per decade logarithmic energy binning.
In particular, the sensitivity curves for \mbox{LST-1} correspond to the average performance below 35\degr\ zenith angle \citep{LSTPerformance}, while the LST array sensitivity curves are computed at 20\degr\protect\footnotemark. 
\footnotetext{Specifically, the IRFs at 20\degr\ in zenith and average azimuth.}
Different novae will be observed with LSTs and will culminate at different zenith angles. Hence, some of them may not be observable at this low zenith angle, implying that the sensitivity at low energies that can be achieved in these observations will degrade. 
When we compares the sensitivity curves for one and four LSTs, the latter will outperform the former by an order of magnitude at energies below $100\,\rm{GeV}$. This improvement is due to the larger collection area and background rejection by the stereo-trigger method. 
Figure~\ref{fig:sensitivity_LSTs} shows that the V906~Car SED model remains below the sensitivity curve of \mbox{LST-1} even for 50 hours of integrated time. Nonetheless, when the CTAO will be operational and the four LSTs dominate the sensitivity of CTAO in the low-energy range, 
the possibility of detecting fainter novae than RS~Oph will grow. 
The LST array sensitivity for 50\,h reaches a flux level similar to the model emission of V906~Car at tens of GeV. This observation time can be challenging to achieve in fast eruptions, but a detection with a shorter integration time may be possible for closer novae than V906~Car. Furthermore, since the nova SED models in Fig.~\ref{fig:sensitivity_LSTs} are time averaged (except RS~Oph), 
the nova spectral models underestimate the emission level in the peak of the gamma-ray phase, when a detection is most probable.

We note, however, that the emission models in Fig.~\ref{fig:sensitivity_LSTs} highly depend on the assumption that the observed spectral shape at HE gamma rays is constrained enough and can be extrapolated to the VHE gamma-ray band. At HE gamma rays, the spectral curvature for most novae is better described by an ECPL shape with cutoff photon energies at a few GeV \citep[e.g.,][]{FermiNovae2014Sci,2016ApJ...826..142C}. Nevertheless, the emissions of RS~Oph and V906~Car, the brightest embedded and classical novae, are better explained by a log-parabola shape \citep{2022_RSOphMAGIC,2022_RSOphHESS,V960Car2020NatAs}. \citet{2022_RSOphMAGIC} suggested that the spectral shape of RS~Oph is not different from other novae, but its bright emission is the cause of the VHE gamma-ray detection. The expected TeV emission in the case of V407~Cyg and V339~Del for an RS Oph-like spectrum would be below the reported UL constraints \citep{2022_RSOphMAGIC}. Under this assumption, the extrapolated emission for the ECPL model for HE novae will underestimate the emission of an RS-Oph-like shape at VHE gamma rays.

Moreover, the VHE emission will depend on the properties of the system and the ejecta if the observed cutoff energy in the photon spectrum of all novae originates from the balance between acceleration and cooling processes. In this regard, \citet{2024arXiv240504469T} explored the capabilities of CTAO to constrain the physical parameters of novae from a modelling and parameter study approach, assuming that the VHE gamma rays are produced through hadronic interactions. The novae detectability and extension in time would depend on the shock velocity and ejecta mass. Furthermore, the parameter space study showed that 30\% of the brightest sample would be detectable with CTAO. These constraints on the VHE emission with multi-wavelength observations will be relevant for constraining the physical parameters of the nova phenomena and for determining whether the physical differences in embedded and classical nova systems also reflect in their gamma-ray emission.

\section{Conclusions}
\label{sect:conclusion}

The source RS~Oph is the first nova that was detected at VHE gamma rays and the first transient source that was detected with the first LST of the future CTAO. During the first observation days after the eruption ($t-t_0 \sim$ 1\,d, 2\,d, and 4\,d), RS~Oph was statistically detected at $6.6\sigma$ with \mbox{LST-1} and was characterised by a soft power-law emission at energies $E=[0.03, 1]\,\mathrm{TeV}$. \mbox{LST-1} spectral results are consistent with the emission reported with MAGIC and \mbox{H.E.S.S.} in coincident observation-day intervals with \mbox{LST-1}. We did not detect RS~Oph with \mbox{LST-1} after 21 days after the eruption onset.

We obtained the particle energy spectrum during the \mbox{LST-1} observations immediately after the eruption onset by using the most complete gamma-ray spectrum to date, including \textit{Fermi}-LAT, \mbox{LST-1}, \mbox{H.E.S.S.}, and MAGIC spectral information. The simpler spectral shape of the hadronic model supports the hadronic over the leptonic scenario to explain the RS~Oph gamma-ray emission, although the relative likelihood of the two models is comparable. The proton energy spectrum evolves with time, increasing the maximum energy of the accelerated protons from hundreds of GeV up to TeV energies between observation-day intervals 1 and 4 after the eruption. The results were validated with a set of robustness tests (Appendix \ref{sect:model_fit_results}).

In the following years, other novae can be expected to be detectable by IACTs. The next eruption of T~CrB is foreseen to occur and likely become a bright nova at gamma-ray energies. An event like this is expected to give outstanding constraints in the evolution of gamma-ray emission during the eruption phase and the maximum energy attainable by the accelerated particles in embedded novae. In the near future, the better sensitivity of the LST array with respect to current facilities at low energies will allow us to probe fainter gamma-ray fluxes. A better sensitivity may enable the detection of classical novae, a nova type yet to be detected at VHE gamma rays.

\begin{acknowledgements}

We thank the anonymous referee for useful suggestions and comments that helped to improve the content of the manuscript. We gratefully acknowledge financial support from the following agencies and organisations:\par

Conselho Nacional de Desenvolvimento Cient\'{\i}fico e Tecnol\'{o}gico (CNPq), Funda\c{c}\~{a}o de Amparo \`{a} Pesquisa do Estado do Rio de Janeiro (FAPERJ), Funda\c{c}\~{a}o de Amparo \`{a} Pesquisa do Estado de S\~{a}o Paulo (FAPESP), Funda\c{c}\~{a}o de Apoio \`{a} Ci\^encia, Tecnologia e Inova\c{c}\~{a}o do Paran\'a - Funda\c{c}\~{a}o Arauc\'aria, Ministry of Science, Technology, Innovations and Communications (MCTIC), Brasil;
Ministry of Education and Science, National RI Roadmap Project DO1-153/28.08.2018, Bulgaria;
Croatian Science Foundation, Rudjer Boskovic Institute, University of Osijek, University of Rijeka, University of Split, Faculty of Electrical Engineering, Mechanical Engineering and Naval Architecture, University of Zagreb, Faculty of Electrical Engineering and Computing, Croatia;
Ministry of Education, Youth and Sports, MEYS  LM2023047, EU/MEYS CZ.02.1.01/0.0/0.0/16\_013/0001403, CZ.02.1.01/0.0/0.0/18\_046/0016007, CZ.02.1.01/0.0/0.0/16\_019/0000754 , CZ.02.01.01/00/22\_008/0004632 and CZ.02.01.01/00/23\_015/0008197 Czech Republic;
CNRS-IN2P3, the French Programme d’investissements d’avenir and the Enigmass Labex, 
This work has been done thanks to the facilities offered by the Univ. Savoie Mont Blanc - CNRS/IN2P3 MUST computing center, France;
Max Planck Society, German Bundesministerium f{\"u}r Bildung und Forschung (Verbundforschung / ErUM), Deutsche Forschungsgemeinschaft (SFBs 876 and 1491), Germany;
Istituto Nazionale di Astrofisica (INAF), Istituto Nazionale di Fisica Nucleare (INFN), Italian Ministry for University and Research (MUR), and the financial support from the European Union -- Next Generation EU under the project IR0000012 - CTA+ (CUP C53C22000430006), announcement N.3264 on 28/12/2021: ``Rafforzamento e creazione di IR nell’ambito del Piano Nazionale di Ripresa e Resilienza (PNRR)'';
ICRR, University of Tokyo, JSPS, MEXT, Japan;
JST SPRING - JPMJSP2108;
Narodowe Centrum Nauki, grant number 2019/34/E/ST9/00224, Poland;
The Spanish groups acknowledge the Spanish Ministry of Science and Innovation and the Spanish Research State Agency (AEI) through the government budget lines PGE2021/28.06.000X.411.01, PGE2022/28.06.000X.411.01 and PGE2022/28.06.000X.711.04, and grants PID2022-139117NB-C44, PID2019-104114RB-C31,  PID2019-107847RB-C44, PID2019-104114RB-C32, PID2019-105510GB-C31, PID2019-104114RB-C33, PID2019-107847RB-C41, PID2019-107847RB-C43, PID2019-107847RB-C42, PID2019-107988GB-C22, PID2021-124581OB-I00, PID2021-125331NB-I00, PID2022-136828NB-C41, PID2022-137810NB-C22, PID2022-138172NB-C41, PID2022-138172NB-C42, PID2022-138172NB-C43, PID2022-139117NB-C41, PID2022-139117NB-C42, PID2022-139117NB-C43, PID2022-139117NB-C44, PID2022-136828NB-C42 funded by the Spanish MCIN/AEI/ 10.13039/501100011033 and “ERDF A way of making Europe; the ``Centro de Excelencia Severo Ochoa" program through grants no. CEX2019-000920-S, CEX2020-001007-S, CEX2021-001131-S; the ``Unidad de Excelencia Mar\'ia de Maeztu" program through grants no. CEX2019-000918-M, CEX2020-001058-M; the ``Ram\'on y Cajal" program through grants RYC2021-032991-I  funded by MICIN/AEI/10.13039/501100011033 and the European Union “NextGenerationEU”/PRTR; RYC2021-032552-I and RYC2020-028639-I; the ``Juan de la Cierva-Incorporaci\'on" program through grant no. IJC2019-040315-I and ``Juan de la Cierva-formaci\'on"' through grant JDC2022-049705-I. They also acknowledge the ``Atracción de Talento" program of Comunidad de Madrid through grant no. 2019-T2/TIC-12900; the project ``Tecnologi\'as avanzadas para la exploracio\'n del universo y sus componentes" (PR47/21 TAU), funded by Comunidad de Madrid, by the Recovery, Transformation and Resilience Plan from the Spanish State, and by NextGenerationEU from the European Union through the Recovery and Resilience Facility; the La Caixa Banking Foundation, grant no. LCF/BQ/PI21/11830030; Junta de Andaluc\'ia under Plan Complementario de I+D+I (Ref. AST22\_0001) and Plan Andaluz de Investigaci\'on, Desarrollo e Innovaci\'on as research group FQM-322; ``Programa Operativo de Crecimiento Inteligente" FEDER 2014-2020 (Ref.~ESFRI-2017-IAC-12), Ministerio de Ciencia e Innovaci\'on, 15\% co-financed by Consejer\'ia de Econom\'ia, Industria, Comercio y Conocimiento del Gobierno de Canarias; the ``CERCA" program and the grants 2021SGR00426 and 2021SGR00679, all funded by the Generalitat de Catalunya; and the European Union's ``Horizon 2020" GA:824064 and NextGenerationEU (PRTR-C17.I1). This research used the computing and storage resources provided by the Port d’Informació Científica (PIC) data center.
State Secretariat for Education, Research and Innovation (SERI) and Swiss National Science Foundation (SNSF), Switzerland;
The research leading to these results has received funding from the European Union's Seventh Framework Programme (FP7/2007-2013) under grant agreements No~262053 and No~317446;
This project is receiving funding from the European Union's Horizon 2020 research and innovation programs under agreement No~676134;
ESCAPE - The European Science Cluster of Astronomy \& Particle Physics ESFRI Research Infrastructures has received funding from the European Union’s Horizon 2020 research and innovation programme under Grant Agreement no. 824064. This work was conducted in the context of the CTA Consortium. This paper has gone through internal review by the CTA Consortium. Author contribution: 
A. Aguasca-Cabot: project coordination, \mbox{LST-1} data analysis, model fitting, paper drafting and edition;
M. I. Bernardos: {\it Fermi}-LAT and \mbox{LST-1} data analysis;
P. Bordas: discussion of model fitting and results;
D. Green: project coordination;
Y. Kobayashi: \mbox{LST-1} data analysis;
R. López-Coto: project coordination;
M. Rib\'o: discussion of model fitting and results;
J. Sitarek: model-fitting codes and edition of the discussion section.
All these authors have participated in the paper drafting. The rest of the authors have contributed in one or several of the following ways: design, construction, maintenance and operation of the instrument(s) used to acquire the data; preparation and/or evaluation of the observation proposals; data acquisition, processing, calibration and/or reduction; production of analysis tools and/or related Monte Carlo simulations; discussion and approval of the contents of the draft.

\end{acknowledgements}

\bibliographystyle{aa}

\bibliography{references}

\begin{appendix}

\section{Daily statistical detection significance}
\label{sect:sig}
The statistical detection significance \citep[formula 17;][]{1983ApJ...272..317L} for the full energy range on each day is 3.2$\sigma$, 2.8$\sigma$ and 5.5$\sigma$ for $t-t_0 =1$\,d, 2\,d and 4\,d, respectively. We note that the event selection cuts used to compute the detection significance are optimised to maximise the LST-1 sensitivity (see definition of sensitivity in Sect.~\ref{sect:discussion_outlook}) under a soft spectrum assumption. Global cuts in gammaness and $\theta$ of 0.9 and 0.1\,deg, respectively, are used in this step. Conversely, the cuts applied to the gamma-like events used to compute the SEDs and light curve are energy dependent. This approach is considered for the latter to reduce the impact of systematic uncertainties. In particular, the mismatch of real/MC data. As a consequence, the energy threshold using energy-dependent and global cuts is different, with the former being lower than the latter in this case. Therefore, the events at the lowest energies of the spectrum, where RS~Oph is bright, do not contribute significantly to the detection significance.

\section{Joint analysis}
\label{sect:joint_analysis}

\begin{figure}[H]
\centering
\includegraphics[width=\hsize]{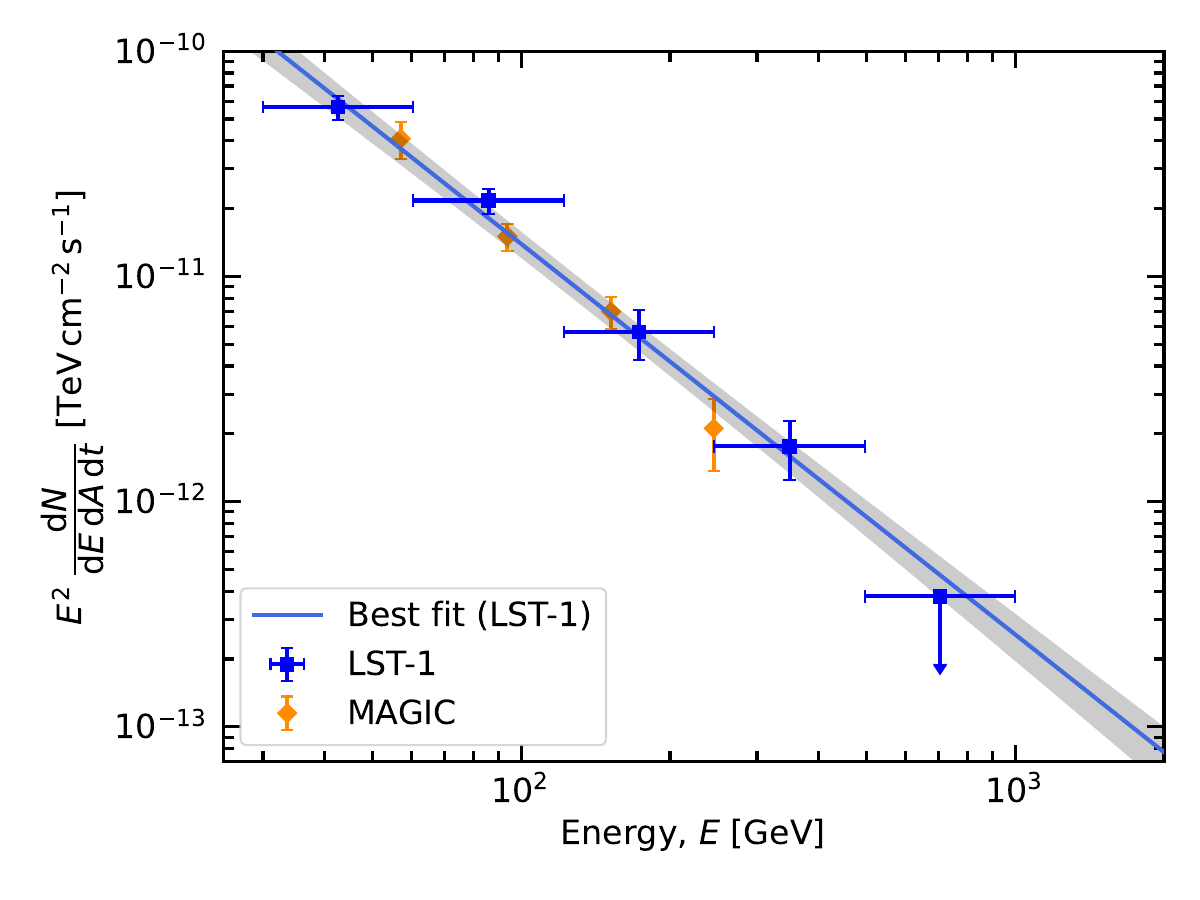}
  \caption{SED at VHE gamma rays obtained with \mbox{LST-1} (blue squares) and MAGIC \citep[orange diamonds;][]{2022_RSOphMAGIC} using all observations between observation-day intervals $t-t_0 \sim[1,4]$\,d (see Table \ref{tab:obs_table} for \mbox{LST-1}). The best-fit model for \mbox{LST-1} is displayed as a blue line together with a grey spectral error band. 
  }
     \label{fig:VHE_joint_SED_LST}  
\end{figure}

The spectral analysis of \mbox{LST-1} data at observation days 1 to 4 altogether is assessed. The \mbox{LST-1} joint spectral analysis follows the approach described in Sect.~\ref{sect:LST1analysis}. The SED for the \mbox{LST-1} observations right after the eruption (observation-day intervals $t-t_0 \sim 1\, \rm{d}, 2\, \rm{d}$ and $ 4\, \rm{d}$, see Table~\ref{tab:obs_table}) is shown in Fig.~\ref{fig:VHE_joint_SED_LST}. The best-fit model parameters for the joint analysis are tabulated in the bottom part of Table~\ref{tab:LST1_best_fit_model}. The best-fit spectral model using all data right after the eruption is soft and compatible with the SED reported by \cite{2022_RSOphMAGIC}, as shown in Fig.~\ref{fig:VHE_joint_SED_LST}. Nonetheless, we note that the MAGIC SED uses data from $t-t_0 \sim 3\, \rm{d}$, while \mbox{LST-1} does not.

\section{Robustness tests of the model fitting}
\label{sect:model_fit_results}

\begin{table}[H]
    \caption{
    Set of input reference parameter values assumed for the hadronic and leptonic model for each observation day.
    }
    \begin{center}
        \footnotesize
        \label{tab:input_params_model}
        \begin{tabular}{lccc}
        \hline
        \hline
        \noalign{\smallskip}
        Parameter  & \multicolumn{3}{c}{Reference value on observation day} \\
        \noalign{\smallskip}
        \cline{2-4}
        \noalign{\smallskip}
           & Day 1 & Day 2 & Day 4 \\
        \noalign{\smallskip}     
        \hline        
        \noalign{\smallskip}
         Slope, $\Gamma_{\rm p}$ & $-$2.2 & $-$2.2 & $-$2.2 \\
         $E_{\rm c,p}$ [$\rm{TeV}$] & $[0.2,3]$ & $[0.2,3]$ & $[0.2,3]$ \\
         Slope 1, $\Gamma_{\rm e,1}$ & $-$0.5 & $-$0.6 & $-$1.5 \\
         Slope 2, $\Gamma_{\rm e,2}$ & ~~$-$3.75 & $-$3.5 & $-$4.0 \\
         $E_{\rm b,e}$ [$\rm{GeV}$] & 15~~ & 10~~ & 35~~ \\
         \mbox{LST-1} syst. & ~~\,0.0 & ~~\,0.0 & ~~\,0.0 \\
         MAGIC syst. & ~~\,0.0 & ~~\,0.0 & ~~\,0.0 \\
         \mbox{H.E.S.S.} syst. & ~~\,0.0 & ~~\,0.0 & ~~\,0.0 \\
        \noalign{\smallskip}
        \hline
        \end{tabular}
    \end{center}
\end{table}

To assess the stability of the fitting process for the hadronic and the leptonic models, the fitting is initialised modifying the initial parameter values of the particle energy distribution around a set of reference values (the input systematic IACT energy-scale values are set to zero for all executions). The reference initial parameters are fixed at the parameter results obtained by \citet{2022_RSOphMAGIC,2022_RSOphHESS}. On the one hand, since the cutoff energy of the proton energy distribution reported by the MAGIC and \mbox{H.E.S.S.} Collaborations are not the same, the input cutoff energy in the fitting is established within the energy range reported by both IACT facilities. On the other hand, the assumed reference input values of the electron energy distribution are set to the best-fit results from MAGIC (see Table~\ref{tab:input_params_model}).

A sequence of five evenly spaced values per parameter (six for the proton cutoff energy to cover the wide input parameter range) is used to define the input parameter space. The range of values we let vary for each free parameter of the particle distribution is within the 50\% relative difference to the corresponding reference input parameters. The reference input parameter values in Table~\ref{tab:input_params_model} are included in the grid. All possible combinations of the particle energy distribution parameters are considered in the set of execution.

The vast majority of the leptonic and hadronic model fitting executions without considering systematics converge (93\% and 92\%, respectively). The former fitting presents more scatter of the fit results than the hadronic one likely due to the larger number of free parameters than the assumed proton spectrum shape. On the contrary, the hadronic model does not present significant variations in the output parameter values for the different input values in the parameter space. The fit results are compatible within errors for the different executions. The spread of the best-fit particle spectrum parameters at the interquartile range (IQR) for the leptonic and hadronic models without allowing for nuisance parameters describing systematics are shown in Table~\ref{tab:parameter_distribution_statistics}. One can see that the slope below the energy break of the electron energy distribution ($\Gamma_{\rm e,1}$) presents the most significant variations. 
The stability of the fitting process for the leptonic model is not as robust as the fitting of the hadronic one. Therefore, the leptonic model using a BPL model cannot constrain the parameters of the electron energy distribution as precisely as the hadronic model with an ECPL.

\begin{table*}
    \caption{
    Median and IQR statistics for the best-fit parameter distributions. 
    }
    \begin{center}
        \footnotesize
        \label{tab:parameter_distribution_statistics}
        \begin{tabular}{llccc}
        \hline
        \hline
        \noalign{\smallskip}

        Model & Parameter & \multicolumn{3}{c}{Median (IQR) on observation day} \\
        \noalign{\smallskip}
        \cline{3-5}
        \noalign{\smallskip}
          & &  Day 1  &  Day 2  &  Day 4 \\
        \noalign{\smallskip}     
        \hline        
        \noalign{\smallskip}
         Leptonic BPL model without systematics & Slope 1, $\Gamma_{\rm e,1}$ & 1.2 (3.5)& $-$0.2 (3.6) & $-$1.3 (0.9) \\
         & Slope 2, $\Gamma_{\rm e,2}$ & $-$3.75 (0.08) & $-$3.4 (0.2) & $-$3.57 (0.07) \\
         & $E_{\rm b,e}$ [$\rm{GeV}$] & 12 (4) & 11 (10) & 24 (11) \\

         \noalign{\smallskip}  
         \noalign{\smallskip}  

         Hadronic ECPL model without systematics & Slope, $\Gamma_{\rm p}$ & $-$2.252 (0.003) & $-$2.4934 (0.0014) & $-$2.476 (0.002) \\
         & $E_{\rm c,p}$ [$\rm{TeV}$] & 0.259 (0.002) &  0.969 (0.008) & 1.64 (0.02) \\

         \noalign{\smallskip}  
         \noalign{\smallskip}          
                          
         Leptonic BPL model with systematics & Slope 1, $\Gamma_{\rm e,1}$ & 0.5 (1.8) & $-$1.4 (1.4) & $-$1.5 (0.8) \\
         & Slope 2, $\Gamma_{\rm e,2}$ & $-$3.76 (0.15) & $-$3.6 (0.3) & $-$3.71 (0.13) \\
         & $E_{\rm b,e}$ [$\rm{GeV}$] & 13 (3) & 18 (14) & 29 (14) \\
         & \mbox{LST-1} syst. & $-$2 (8) & $-$7 (9) & $-$1 (8) \\
         & MAGIC syst. & 1 (9) & 1 (10) & 9 (8) \\
         & \mbox{H.E.S.S.} syst. & 4 (10) & $-$8 (11) & $-$12 (4) \\

         \noalign{\smallskip}  
         \noalign{\smallskip}  
         
         Hadronic ECPL model with systematics & Slope, $\Gamma_{\rm p}$ & $-$2.18 (0.12) & $-$2.45 (0.07) & $-$2.40 (0.05) \\
         & $E_{\rm c,p}$ [$\rm{TeV}$] & 0.22 (0.08) & 0.7 (0.2) & 1.0 (0.2) \\
         & \mbox{LST-1} syst. & 4 (5) & $-$4 (9) & $-$3 (10) \\
         & MAGIC syst. & 0 (6) & 7 (7) & 10 (7) \\
         & \mbox{H.E.S.S.} syst. & $-$1 (8) & $-$7 (8) & $-$10 (5) \\

        \noalign{\smallskip}
        \hline
        \end{tabular}
    \end{center}
\end{table*}

\begin{figure}
\centering
\includegraphics[width=\hsize]{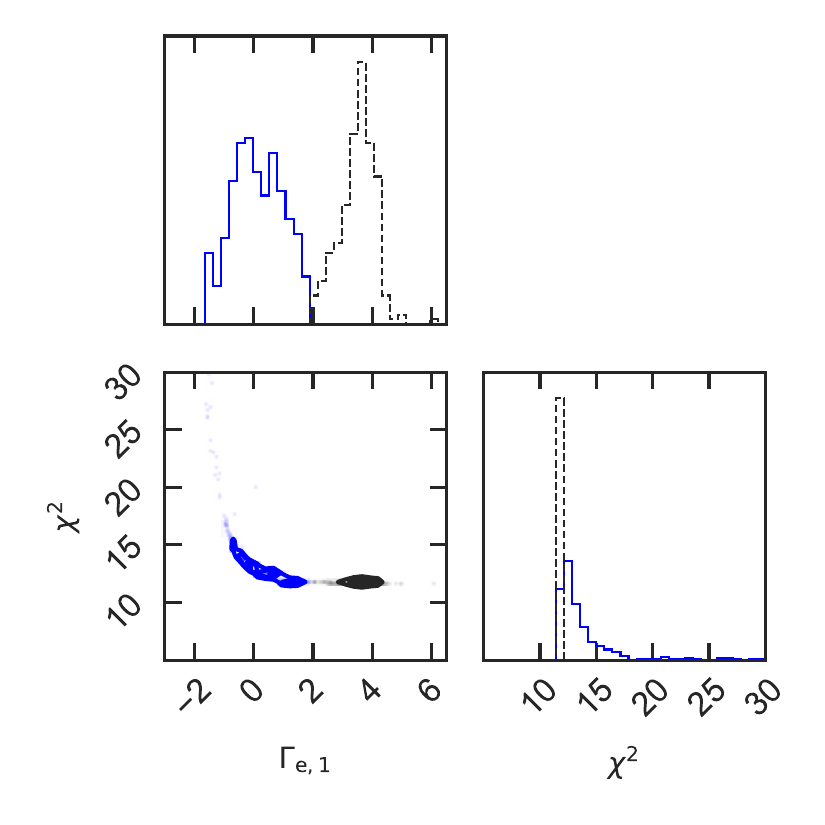}
  \caption{Corner plot of the best-fit $\Gamma_{\rm e,1}$ for observation day 1 and the associated $\chi^2$ fit value for different input values in the parameter space. The best-fit models used to provide the leptonic results without systematics in Table~\ref{tab:model_fitting_results} are shown as solid blue lines, while the ones excluded are shown as dashed black lines (see text).}
     \label{fig:slope_el_1}
\end{figure}

Most of the best-fit parameter values assemble forming unimodal distributions for the leptonic and hadronic models. The inclusion of \mbox{LST-1} into the dataset reduces the dispersion of the distribution of the best-fit parameters. Nevertheless, two local minima are visible in the output parameter distributions for the leptonic fitting of observation day 1 ($t-t_0 \sim1\,\textrm{d}$), hence a bimodal distribution (see Fig.~\ref{fig:slope_el_1}, where the two distributions are shown in blue and black). $\Gamma_{\rm e,1}$ presents a local minimum peaking at negative values close to zero, while the second minimum is centred at positive values. A bump towards the positive local minimum values is also noticeable for observation day 2 ($t-t_0 \sim2\,\textrm{d}$). The two local minima are smoothly connected in the $\chi^2$ space. The local minimum close to zero peaks at a $\chi^2$ value of $\chi^2=12.1$, while the other minimum is centred at $\chi^2=11.7$. The two local minima are possible solutions for fitting the curvature of the HE spectrum. Similarly, a bimodal distribution is also obtained if we execute the fitting of the model published by \citet{2022_RSOphMAGIC} together with the \textit{Fermi}-LAT and MAGIC data as input spectral information, yet the fitting in \citet{2022_RSOphMAGIC} restricted the fitting range to $\Gamma_{\rm e,1}<-0.5$.

Although the solutions with a hard electron spectrum are difficult to explain through the classical non-relativistic DSA mechanism, we extend the parameter space of the model fitting to include such solutions to study the stability of the model and provide reliable uncertainties. We report the results excluding the best-fit models with $\Gamma_{\rm e,1}>2$ (black distribution in Fig.~\ref{fig:slope_el_1}), which is the intersection value we obtain after fitting two normal distributions to the distribution of the best-fit $\Gamma_{\rm e,1}$ results for observation day 1 with different input parameter values (see Fig.~\ref{fig:slope_el_1}). 

On the one hand, restricting the results to $\Gamma_{\rm e,1}<2$ does not significantly impact the results presented in this work for several reasons. Firstly, the best-fit slope above the energy break would not present a notable change because it is constrained by the IACT data. Whilst a slight decrease of the energy break value would result from allowing the excluded range of $\Gamma_{\rm e,1}$. The minor differences in these parameters would be compatible within uncertainties to the reported value in Table~\ref{tab:model_fitting_results}. Secondly, the physical interpretation of the excluded solutions is the same as the ones considered in the valid range (see Sect.~\ref{sect:discussion_outlook}). Thirdly, the improvement in the fit statistics of including such solutions is marginal. Therefore, no preference between the leptonic and hadronic models could be drawn with the AIC estimator when including these solutions. On the other hand, if the fitting range of the electron spectral slope below the energy break would had been restricted to values $\Gamma_{\rm e,1}<-0.5$, as in \citet{2022_RSOphMAGIC}, the goodness of fit of the leptonic model would have been worse than most of the output models obtained in this work (see Fig.~\ref{fig:slope_el_1}). This is attributed to the leptonic model inability to account for the positive gamma-ray emission slope below $1\,\textrm{GeV}$ and posterior decay at higher energies clearly visible on observation day 1. Consequently, constraining the fitting range would have favoured the hadronic model. However, such change does not affect the fit results for observation days 2 and 4 because the emission with \textit{Fermi}-LAT is dimmer and the curvature at HE gamma rays is less pronounced ($\textrm{p-value}=0.09$ and $0.83$ under the assumption of a flat HE emission for observation days 2 and 4, respectively). As a result, the best-fit leptonic model does not require a strong spectral slope break as in observation day 1 to provide a good fit. Then, both the leptonic and the hadronic models achieve comparable goodness of fit. In this manner, a nova with a brighter gamma-ray emission than RS~Oph, e.g., \mbox{T~CrB}, should help constrain the population of relativistic particles responsible for the gamma-ray emission.

The model fit results in Table~\ref{tab:model_fitting_results} correspond to the best-fit execution result that is the closest one to the middle of both the AIC distribution and the particle spectrum parameter distributions obtained when varying the input value. If the energy-scale systematic distributions are considered, they are used as well. We report the uncertainties of the best-fit model parameters results as the quadratic sum of the parameter uncertainty of the fitting and the IQR of the output parameters for the different input value executions.

Despite the scattered fit results for $\Gamma_{\rm e,1}$, the slope above the energy break ($\Gamma_{\rm e,2}$) is well constrained for all days thanks to the IACT spectral information at VHE gamma rays. The use of the \mbox{LST-1} spectral information is relevant to constrain the parameter space of solutions, which is narrower with the \mbox{LST-1} data because of the smaller energy gap between \textit{Fermi}-LAT and the IACTs data if the \mbox{LST-1} SED points are considered.

\section{Model fitting with systematics}
\label{sect:fitting_with_syst}
When allowing for energy-scale systematics as nuisance parameters in the model fitting, the percentage of executions that converge drops to 79\% and 81\% for the leptonic and hadronic models, respectively, with respect to the modelling without energy-scale systematics.
The output parameter distribution statistics for the modelling with systematics are shown in Table~\ref{tab:parameter_distribution_statistics}. One can notice that the parameter distribution IQRs are wider when systematics are considered than when they are not. The broader distributions are expected due to the degeneracy of adding extra parameters in the model fitting to allow shifts with respect to the original SED points. Moreover, IQR values are higher for the leptonic model than the hadronic model (expected due to the worse stability of the leptonic model than the hadronic one). The local minimum peaking at positive $\Gamma_{\rm e,1}$ value present in the leptonic modelling without systematics vanishes when systematics are considered (see Appendix~\ref{sect:model_fit_results}). However, a long tail towards positive values remains.

\begin{figure}
\centering
\includegraphics[width=\hsize]{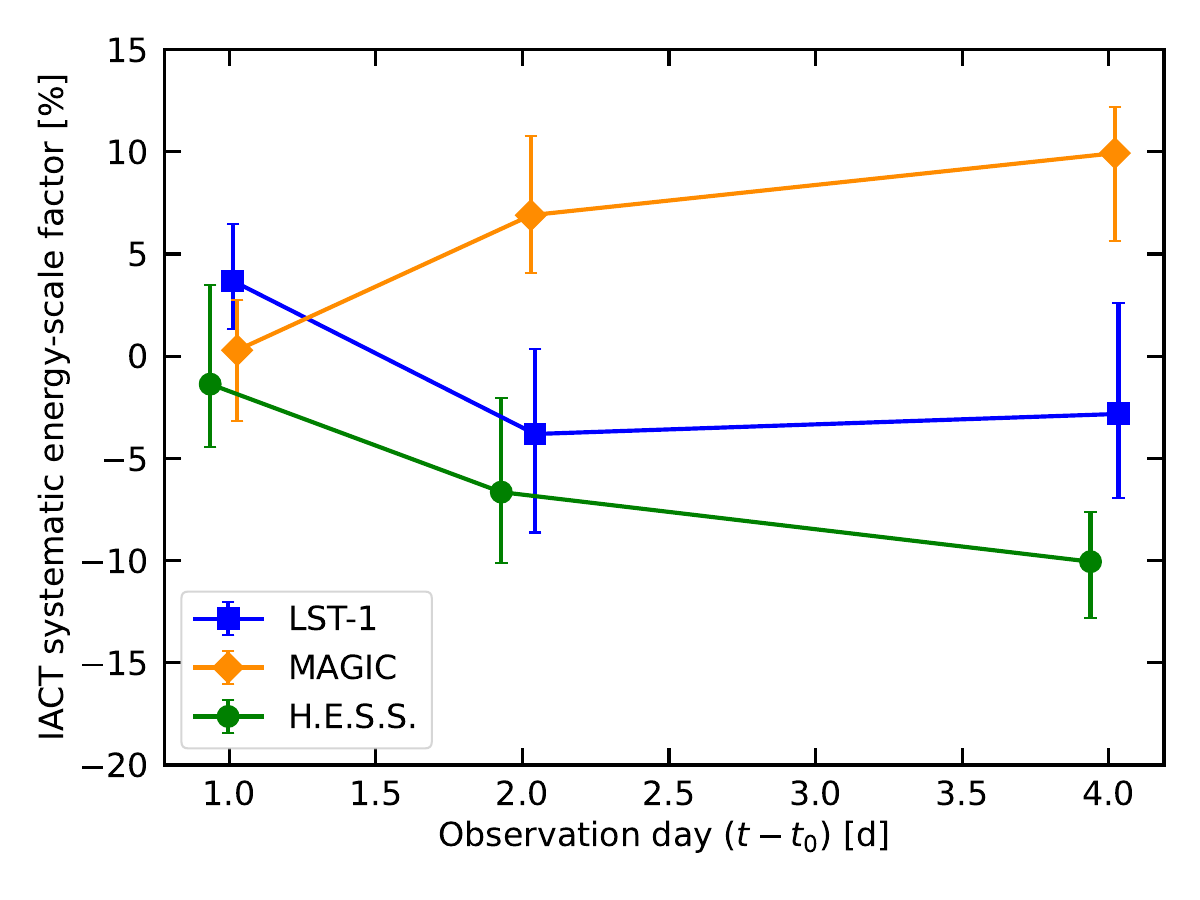}
  \caption{Derived systematic energy scaling factors for \mbox{LST-1} (blue squares), MAGIC (orange diamonds) and \mbox{H.E.S.S.} (green circles) for the different observation-day intervals ($t-t_0 \sim$ 1\,d, 2\,d and 4\,d). The IACT systematic energy-scale value corresponds to the median value of the IACT distribution obtained with the hadronic model varying the input parameters. Error bars correspond to the IQR.}
     \label{fig:IACT_syst_evo}
\end{figure}

The distributions of the best-fit IACTs energy-scale values for different input parameters form unimodal distributions. The spread of the distributions ranges all the allowed values ($\pm 15\%$) for most of the cases. Interestingly, the sign and the time evolution of the peak of the derived systematic energy-scale factors for MAGIC and \mbox{H.E.S.S.} agree with the mismatch between the spectrum from MAGIC and \mbox{H.E.S.S.}, and the fact that the VHE gamma-ray emission increases in time, making the difference more noticeable. Figure~\ref{fig:IACT_syst_evo} shows the best-fit IACT systematic energy scaling factors obtained with the hadronic model for the different observation days. During observation days 1, 2 and 4, the best-daily-fit systematic energy-scale values remain below 5\% for \mbox{LST-1}, while the sequence of MAGIC and \mbox{H.E.S.S.} energy scaling factors start with a low-systematic value for observation day 1 and monotonically increase in time reaching an absolute systematic energy-scale value of about 10\%. MAGIC systematic values remain positive for all the observation days, while \mbox{H.E.S.S.} ones are negative (see Fig.~\ref{fig:IACT_syst_evo}). As expected, no correlation between the best-fit IACT systematic uncertainties is observed.

\begin{figure}[h!]
        \centering
        \includegraphics[clip,width=1\columnwidth]{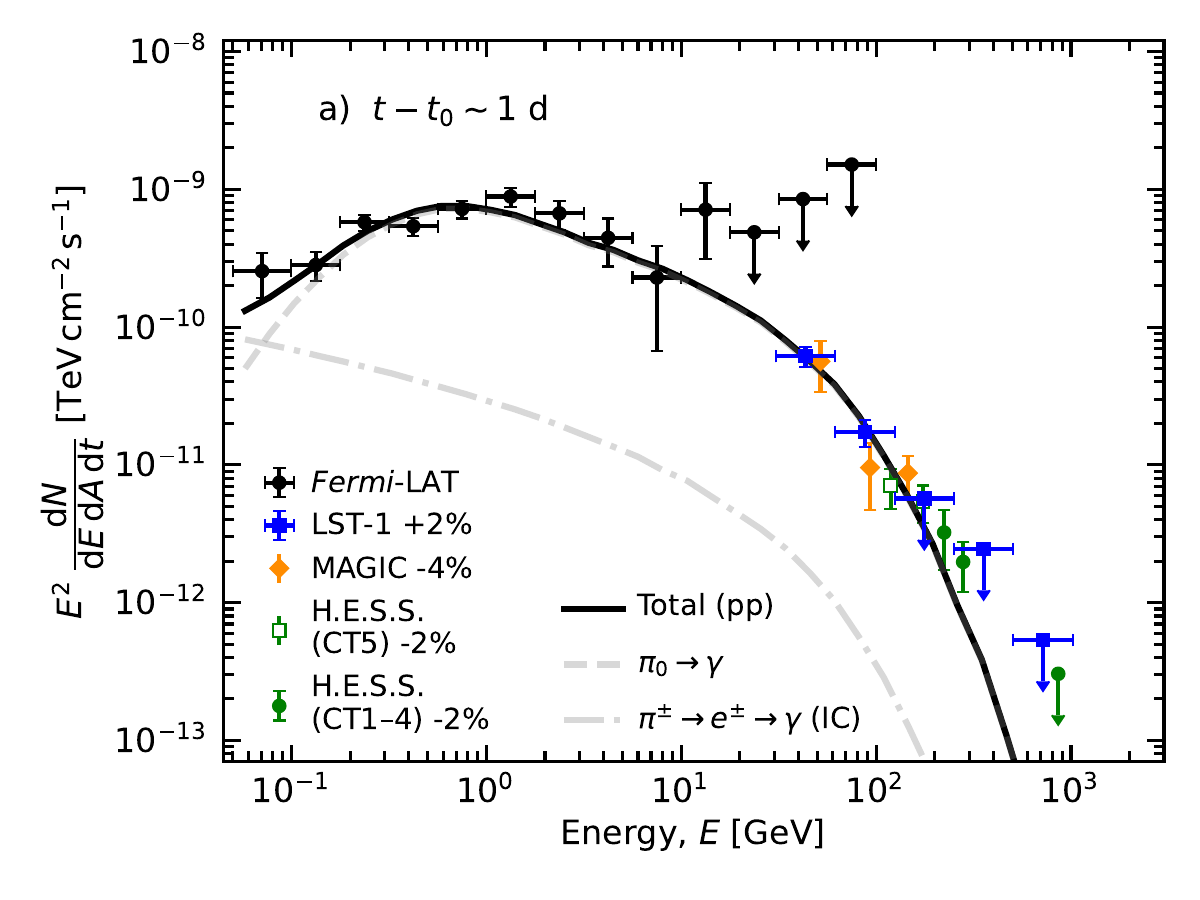}
        
        \includegraphics[clip,width=1\columnwidth]{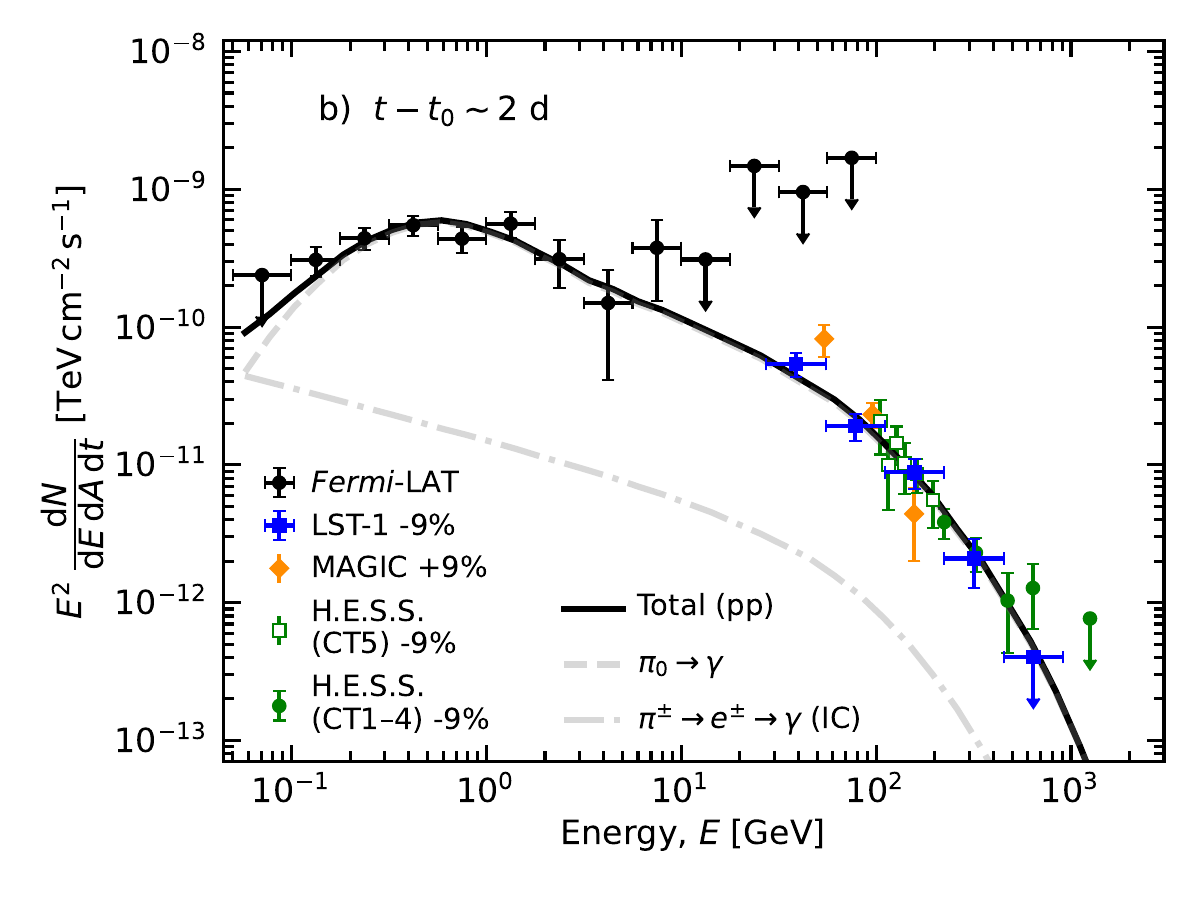}
        
        \includegraphics[clip,width=1\columnwidth]{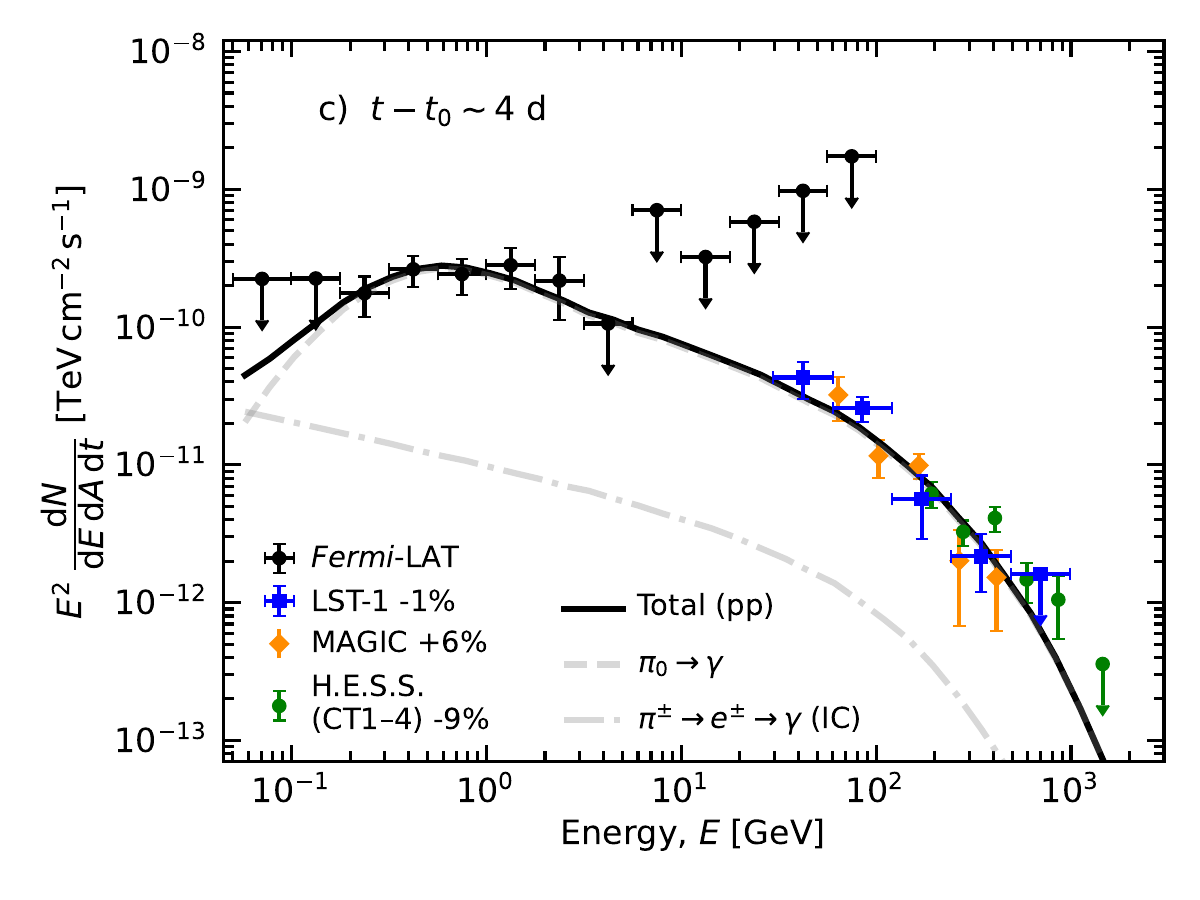}
        \caption{Same SEDs as Fig.~\ref{fig:daily_modelling_FermiLST-MAGIC-HESS} but with the IACT SED points scaled to account for the systematic energy scaling factors obtained for the best-fit hadronic model with systematics. The systematic value for each IACT is shown in the legend. In addition, the best-fit hadronic model considering energy-scale systematics is displayed as a black curve. The corresponding contributions from neutral and charged-pion decays are shown in grey.}
            \label{fig:daily_modelling_FermiLST-MAGIC-HESS_syst}   
\end{figure}

The addition of systematic uncertainties in the energy scale causes the differential fluxes of the IACTs to shift accordingly in the SED, both in the x- and y-axis, with respect to the original ones when the systematic energy scaling factor is different from zero. The daily SEDs adopting the best-fit systematic energy-scale factors in the IACT SED data points are shown in Fig.~\ref{fig:daily_modelling_FermiLST-MAGIC-HESS_syst}. One can see that the differential fluxes at VHE gamma rays are concentrated in a band with less spread than the one displayed in Fig.~\ref{fig:daily_modelling_FermiLST-MAGIC-HESS} without systematic uncertainties taken into account. We note that the best-fit values of the IACT systematics in Fig.~\ref{fig:daily_modelling_FermiLST-MAGIC-HESS_syst} are not the same as the centre values of the IACT systematics in Fig.~\ref{fig:IACT_syst_evo} because the latter shows the median value and the IQR as error bars, while the former is the closest execution result with respect to both the median AIC and particle parameter distributions (see Appendix~\ref{sect:model_fit_results}).

Although the hadronic model with systematics effectively describes the gamma-ray emission, due to rather large statistical uncertainties on the SED data points, taking into account systematic uncertainties as extra fitting parameters (three in this case) does not impact the fit statistics enough to obtain a better fit given the reduction of degrees of freedom. Yet, systematics are foreseen to affect the SED data points. Moreover, it is worth noting that the same systematic energy-scale factor is assumed for the \mbox{H.E.S.S.} spectral information obtained with the array of four telescopes (CT1--4; stereo analysis) and the fifth telescope (CT5; monoscopic analysis) despite not following the exact same analysis. However, the fine agreement between both supports our simplification to reduce the number of parameters in the model fitting.

\section{Lepto-hadronic model}
\label{sect:lephad}

The modelling of the RS~Oph gamma-ray emission using two populations of relativistic particles, electrons and protons, is also studied. The lepto-hadronic model in \citet{2022_RSOphMAGIC} is considered. The particle energy distribution is assumed to follow an ECPL model, for which the cutoff energies of protons and leptons are connected by the acceleration and cooling balance \citep{2022_RSOphMAGIC}. Since including systematic energy-scale uncertainties as nuisance parameters are not preferred for the leptonic and hadronic models, they are excluded in the lepto-hadronic model fitting process to reduce the complexity of the fitting. The fit results are shown in Table~\ref{tab:model_fitting_lephad_results} and the best-fit model emission is shown on the daily SEDs in Fig.~\ref{fig:daily_modelling_FermiLST-MAGIC-HESS_lepthad}. The same procedure as in Appendix~\ref{sect:model_fit_results} is followed to provide the uncertainties on the parameter values.

\begin{table}[h!]
    \caption{ Model-fit results of the lepto-hadronic modelling for observation days $t-t_0 \sim1$\,d, 2\,d and 4\,d. 
    }
    \begin{center}
        \footnotesize
        \label{tab:model_fitting_lephad_results}
        \begin{tabular}{lccc}
        \hline
        \hline
        \noalign{\smallskip}
        Parameter  & \multicolumn{3}{c}{Best-fit value on observation day} \\
        \cline{2-4}
                \noalign{\smallskip}

                   &  Day 1  &  Day 2  &  Day 4 \\
        \noalign{\smallskip}
        \hline
        \noalign{\smallskip}

        \multicolumn{4}{c}{Lepo-hadronic model} \\
        \noalign{\smallskip}
        \hline
        \noalign{\smallskip}
        Slope, $\Gamma$ & $-2.25^{+0.17}_{-0.17}$~~~ & $-1.4^{+0.2}_{-0.6}$~~~ & $-2.49^{+0.05}_{-0.06}$~~~ \\
        \noalign{\smallskip}
        $E_{\rm c, e}$ [$\rm{GeV}$] & $1.1^{+0.6}_{-0.6}$ &  $26^{+6}_{-5}$ & $290^{+90}_{-80}$ \\    
        \noalign{\smallskip}
        $E_{\rm c, p}$ [$\rm{TeV}$] & $0.26^{+0.14}_{-0.14}$ &  $1.1^{+0.3}_{-0.2}$ & $1.8^{+0.5}_{-0.5}$ \\    
        \noalign{\smallskip}        
        $\rm{L}_{\rm p}/\rm{L}_{\rm e}$ [\%] & $940^{+2480}_{-660}$ &  $64^{+20}_{-7}$ & $270^{+1500}_{-270}$ \\    
        \noalign{\smallskip}                
        $\chi^2/N_{\rm d.o.f}$ & $21.2/14$ & $23.1/21$ & $26.9/15$ \\
        \noalign{\smallskip}        
        $\chi^2_{\rm red}$ & $~~~~1.51$ & $~~~~1.10$ & $~~1.79$ \\        
        \noalign{\smallskip}
        AIC$_{\rm c}$ & $32.1$ & $33.1$ & $37.8$ \\
        \noalign{\smallskip}                
        \hline        
        \end{tabular}
    \end{center}
    \tablefoot{
    $\Gamma$ is the best-fit slope for the electron and proton energy distributions. The best-fit cutoff energy of protons and electrons is $E_{\rm c,p}$ and $E_{\rm c,e}$, respectively. We note that the cutoff energies of protons and electrons are in TeV and GeV, respectively. Both spectral energy distributions are related using the proton-to-electron luminosity ratio, $\rm{L}_{\rm p}/\rm{L}_{\rm e}$. We provide the $\chi^2_{\rm red}$ fit statistics ($\chi^2_{\rm red}=\chi^2/N_{\rm d.o.f}$) and the daily AIC$_{\rm c}$ values. The sum of the AIC$_{\rm c}$ values for all days for the lepto-hadronic model is 102.9.}
\end{table}

\begin{figure}[h!]
        \centering

        \includegraphics[clip,width=1\columnwidth]{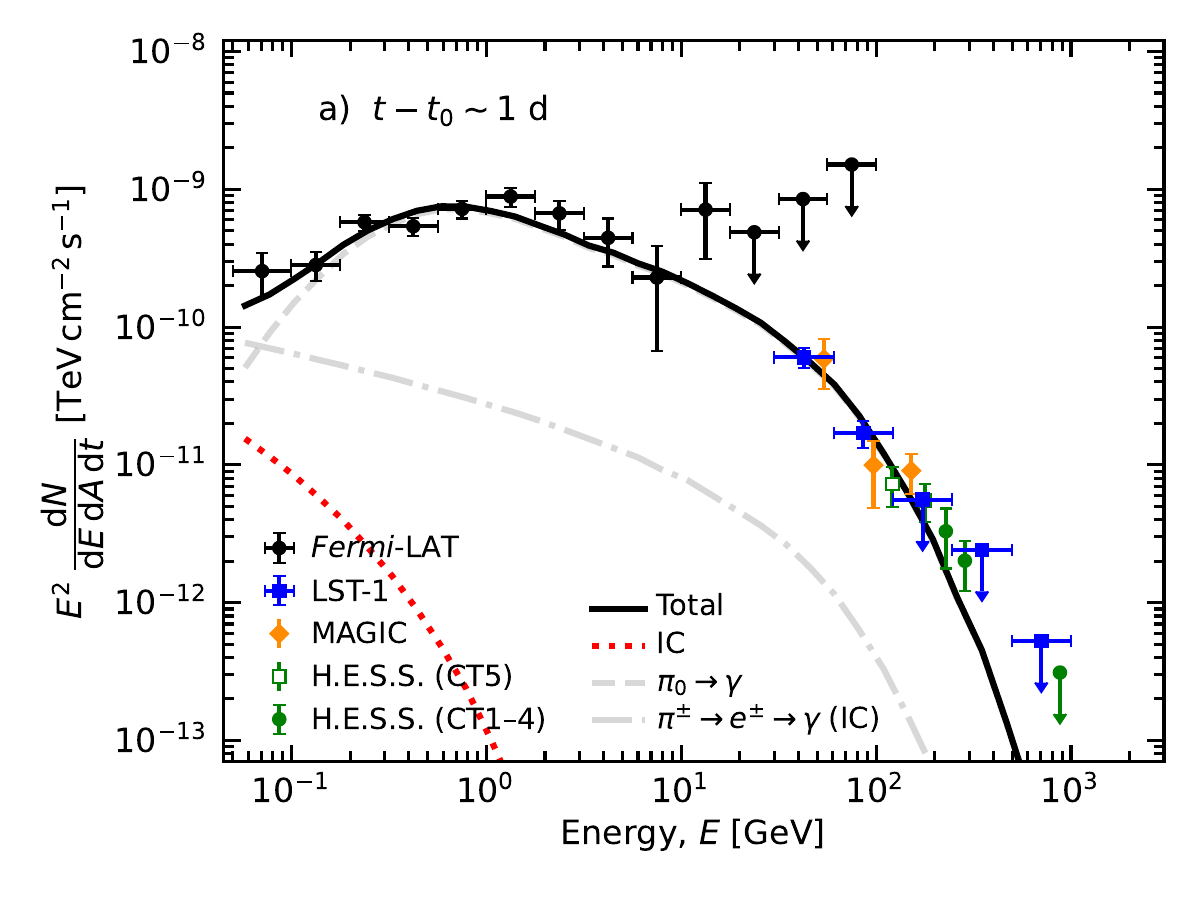}
        
        \includegraphics[clip,width=1\columnwidth]{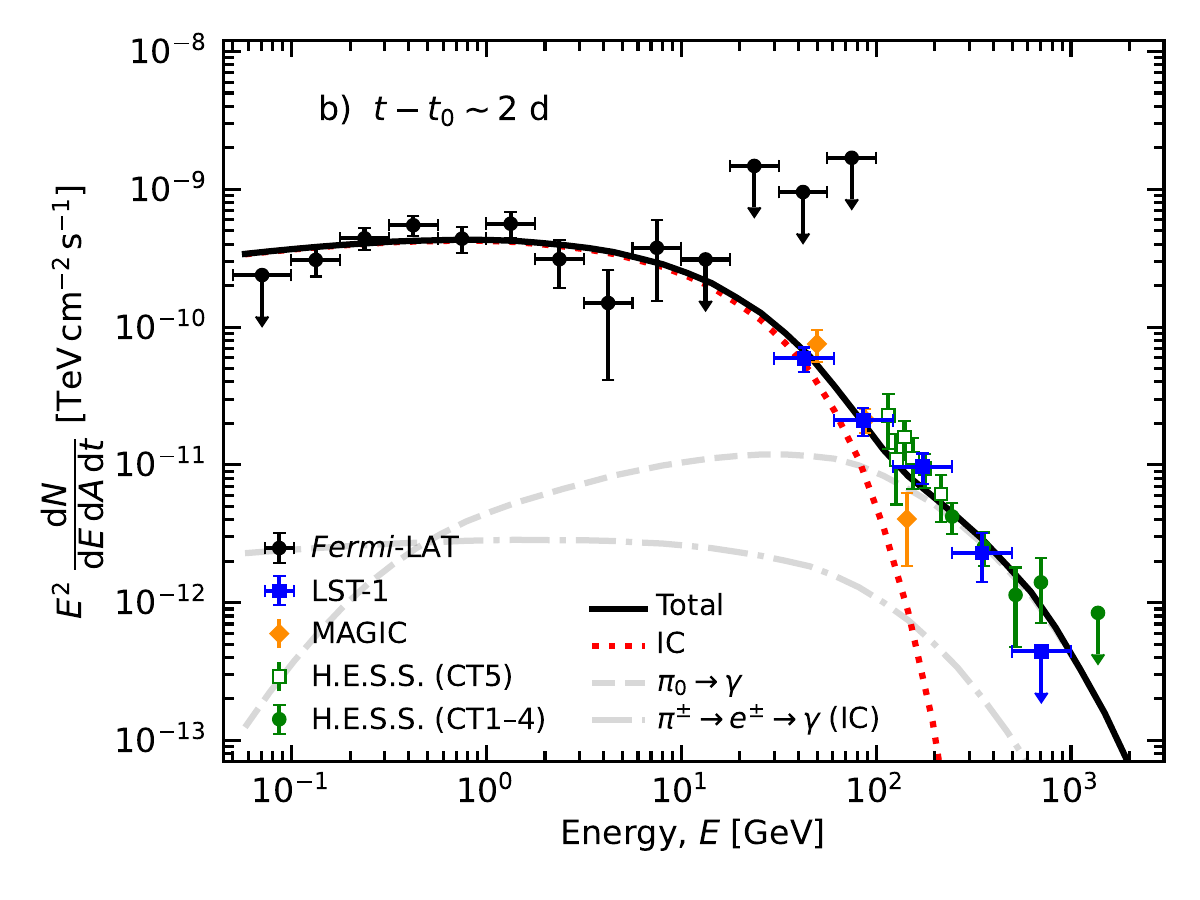}
        
        \includegraphics[clip,width=1\columnwidth]{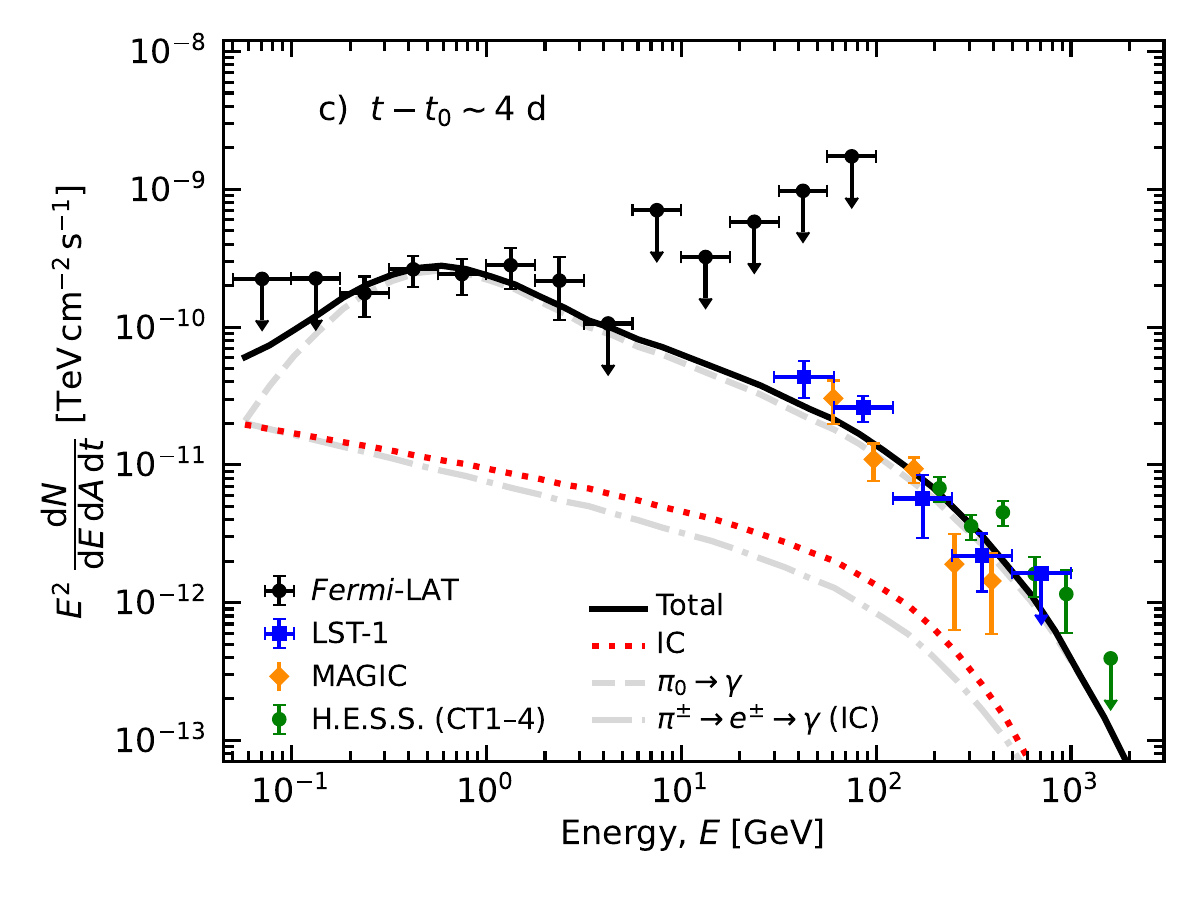}

        \caption{
        Same SEDs as Fig.~\ref{fig:daily_modelling_FermiLST-MAGIC-HESS} but for the best-fit lepto-hadronic model, whose total contribution is displayed as a black curve. The corresponding contributions from neutral and charged-pion decays are shown in grey and the leptonic contribution in red.        
        }
            \label{fig:daily_modelling_FermiLST-MAGIC-HESS_lepthad}        
\end{figure}

The emission of the best-fit lepto-hadronic model on observation days 1 and 4 is dominated by proton-proton interactions in both HE and VHE bands with the spectral index and cutoff energies of the proton energy distribution consistent with the values obtained with the hadronic model (see Sect.~\ref{sect:modelling_results}). On the contrary, the emission at HE gamma rays on observation day 2 originates from IC losses, while the VHE component comes from proton-proton interactions. However, the best-fit model exceeds the emission constrained by the UL in the 50--100\,MeV energy bin, which is important to constrain the curvature of the spectrum. For all days, the Lp/Le ratio remains high, the lowest Lp/Le ratio is $64^{+20}_{-7}\,\%$ on $t-t_0 \sim2\,\textrm{d}$ (see Table~\ref{tab:model_fitting_lephad_results}). We note that the best-fit model is highly dependent on the input parameters. The leptonic component can contribute to the HE emission for some solutions on $t-t_0 \sim1\,\textrm{d}$, but with worse fit statistics and the VHE emission from proton-proton interactions above the SED ULs provided by the IACTs.

When comparing the AIC$_{\rm c}$ values of the lepto-hadronic model with the leptonic and hadronic models, the lepto-hadronic model is less preferred over them (relative likelihood of 0.014 and 0.02, respectively). Given the no preference of the lepto-hadronic model over the hadronic/leptonic model and proton-proton interactions dominate the best-fit model on $t-t_0 \sim1\,\textrm{d}$ and $4\,\textrm{d}$, we disfavour this model. However, the presence of additional components in the initial gamma-ray emission cannot be excluded, but exploring this possibility would exceed the scope of this study and would require additional observations along with detailed multiwavelength modelling.

\section{Akaike weights of the models}
\label{sect:AICweights}
For a better interpretation of the relative likelihoods with respect to the best model according to the AIC, they are normalised to the set of positive weights that sum to 1, called Akaike weights. The information criteria measures for the different models are presented in Table~\ref{tab:AICscores}.

\begin{table}[h!]
    \caption{ Information criteria measures for the models used in this work. 
    }
    \begin{center}
        \footnotesize
        \label{tab:AICscores}
        \begin{tabular}{lccc}
        \hline
        \hline
        \noalign{\smallskip}        
        Model & $\Delta$AIC$_{\rm c}$ & Relative likelihood & Akaike weight \\
        \noalign{\smallskip}
        \hline
        \noalign{\smallskip}
        \makecell[l]{Leptonic BPL \\ w/o systematics} & 0.0 & 1 & 0.58 \\
        \noalign{\smallskip}
        \makecell[l]{Hadronic ECPL \\ w/o systematics} & 0.83 & 0.66 & 0.38 \\
        \noalign{\smallskip}
        \makecell[l]{Leptonic BPL \\ w/ systematics} & 27.04 & 1.35$\times 10^{-6}$ & 7.75$\times 10^{-7}$ \\
        \noalign{\smallskip}
        \makecell[l]{Hadronic ECPL \\ w/ systematics} & 21.65 & 2$\times 10^{-5}$ & 1.14$\times 10^{-5}$\\
        \noalign{\smallskip}
        Lepto-hadronic ECPL & 8.5 & 0.014 & 8.52$\times 10^{-3}$ \\
        \noalign{\smallskip}                
        \hline        
        \end{tabular}
    \end{center}
    \tablefoot{ The $\Delta$AIC$_{\rm c}$ and the relative likelihood values are shown for the leptonic BPL without (w/o) and with (w/) systematics, hadronic ECPL w/o and w/ systematics and the lepto-hadronic ECPL models after summing all days. The $\Delta$AIC$_{\rm c}$ and the relative likelihood values are computed with respect to the model with $\sum$AIC$_{\rm c, min}$. Also, the Akaike weights (see text) are presented.}
\end{table}

\end{appendix}

\end{document}